\documentclass{article}

\pdfoutput=1

\usepackage{arxiv}

\usepackage[utf8]{inputenc} % allow utf-8 input
\usepackage[T1]{fontenc}    % use 8-bit T1 fonts
\usepackage{url}            % simple URL typesetting
\usepackage{booktabs}       % professional-quality tables
\usepackage{amsfonts}       % blackboard math symbols
\usepackage{nicefrac}       % compact symbols for 1/2, etc.
\usepackage{microtype}      % microtypography
\usepackage{amsmath}
\usepackage{graphicx}
\usepackage{doi}
\usepackage{color}
\usepackage{soul}
\usepackage{makecell}
\usepackage{listings, xcolor}
\usepackage{authblk}        % Multiple affiliations variant of author block
\usepackage{caption}
\usepackage{dirtytalk}
\usepackage{tabularx}
\usepackage{booktabs}

\lstdefinelanguage{TOML}{
    comment = [l]{\#},
    keywords = {true, false},
    morestring = [b]{"},
    morekeywords = {open, scale, naics, fips, schedule, school, work, social, travel, attendance}
}

\definecolor{codegreen}{rgb}{0,0.6,0}
\definecolor{codegray}{rgb}{0.5,0.5,0.5}
\definecolor{codepurple}{rgb}{0.58,0,0.82}
\definecolor{backcolour}{rgb}{0.95,0.95,0.95}

\lstdefinestyle{codestyle}{
    backgroundcolor=\color{backcolour},   
    commentstyle=\color{codegreen},
    keywordstyle=\color{green!45!blue},
    numberstyle=\tiny\color{codegray},
    stringstyle=\color{codepurple},
    basicstyle=\ttfamily\footnotesize,
    breakatwhitespace=false,         
    breaklines=true,                 
    captionpos=b,                    
    keepspaces=true,                 
    numbers=left,                    
    numbersep=5pt,                  
    showspaces=false,                
    showstringspaces=false,
    showtabs=false,                  
    tabsize=2
}

\newcommand{\alttilde}{\raise.17ex\hbox{$\scriptstyle\mathtt{\sim}$}}

\title{Epicast 2.0: a large-scale, demographically detailed, agent-based model for simulating respiratory pathogen spread in the United States}

\date{}

\author[1,2]{
	Prescott C.~Alexander
}
\author[1,3]{
	Thomas J.~Harris
}
\author[2,4]{
	Joy~Kitson
}
\author[5]{
	Joseph V.~Tuccillo
}
\author[6]{
	Sara Y.~Del Valle\thanks{\texttt{sdelvall@lanl.gov}}
}
\author[2]{
	Timothy C.~Germann\thanks{\texttt{tcg@lanl.gov}}
}
\vspace{10pt}
\affil[1]{Information Systems and Modeling group (A-1), Los Alamos National Laboratory, Los Alamos, NM, USA}
\affil[2]{Physics and Chemistry of Materials group (T-1), Los Alamos National Laboratory, Los Alamos, NM, USA}
\affil[3]{School of Computing and Information Systems, University of Melbourne, Parkville, Victoria, AUSTRALIA}
\affil[4]{Department of Computer Science, University of Maryland, College Park, MD, USA}
\affil[5]{Geospatial Science and Human Security Division, Oak Ridge National Laboratory, Oak Ridge, TN, USA}
\affil[6]{Associate Laboratory Directorate for Global Security, Los Alamos National Laboratory, Los Alamos, NM, USA}

\renewcommand{\shorttitle}{Epicast 2.0}

\hypersetup{
pdftitle={Epiacst 2.0},
pdfsubject={q-bio.PE, q-bio.QM, physics.soc-ph},
pdfauthor={Prescott C.~Alexander, et al.},
pdfkeywords={Agent-based modeling, computational epidemiology, infectious disease},
}

\begin{document}
\maketitle

\begin{abstract}
	The recent history of respiratory pathogen epidemics, including those caused by influenza and SARS-CoV-2, has highlighted the urgent need for advanced modeling approaches that can accurately capture heterogeneous disease dynamics and outcomes at the national scale, thereby enhancing the effectiveness of resource allocation and decision-making. In this paper, we describe Epicast 2.0, an agent-based model that utilizes a highly detailed, synthetic population and high-performance computing techniques to simulate respiratory pathogen transmission across the entire United States. This model replicates the contact patterns of over 320 million agents as they engage in daily activities at school, work, and within their communities. Epicast 2.0 supports vaccination and an array of non-pharmaceutical interventions that can be promoted or relaxed via highly granular, user specified policies. We illustrate the model's capabilities using a wide range of outbreak scenarios, highlighting the model's varied dynamics as well as its extensive support for policy exploration. This model provides a robust platform for conducting \emph{what if} scenario analysis and providing insights into potential strategies for mitigating the impacts of infectious diseases.
\end{abstract}


\section{Introduction}

Infectious diseases remain a persistent and evolving threat to global health security, contributing to millions of deaths and significant economic disruption each year, as demonstrated by recent, large-scale outbreaks caused by SARS-CoV-2 and influenza. In response to the urgent need for informed decision-making, mathematical and computational models have become critical tools for understanding disease dynamics, evaluating interventions, and guiding policy during outbreaks.

The COVID-19 pandemic led to a particularly large increase in modeling efforts globally, with diverse approaches informing real-time decision-making \cite{rayEnsembleForecastsCoronavirus2020, walkerImpactCOVID19Strategies2020, cramerUnitedStatesCOVID192022, howertonEvaluationUSCOVID192023}. These models span a spectrum of complexity, from simple compartmental models like Susceptible-Infectious-Recovered (SIR) that assume homogeneous mixing, to metapopulation models incorporating spatial structure, to agent-based models (ABMs) that simulate the behavior and interactions of individuals in fine-grained social and geographic environments.

ABMs have become especially valuable in recent years due their ability to simulate heterogeneity in behavior, contact patterns, and intervention effects. By representing individuals as autonomous agents with attributes such as age, location, and occupation, ABMs can assess complex intervention strategies, such as targeted school closures, vaccination rollouts, or work-from-home policies, that are difficult to evaluate with aggregate models \cite{tracyAgentBasedModelingPublic2018, steinhofelAgentBasedModelsVirus2025}.

Building on this need for realistic, high-resolution epidemic simulations, we present Epicast 2.0, an enhanced version of the Epicast agent-based model. Epicast has been previously used in the context of pandemic preparedness and response within the United States, including evaluations of mitigation strategies and school-based interventions \cite{germannMitigationStrategiesPandemic2006,germannSchoolDismissalPandemic2019,delvalleEpiCastSimulatingEpidemics2021,germannAssessingK12School2022}. Epicast 2.0 incorporates updated demographic and mobility data, improved behavioral modeling, and a scalable software architecture designed to simulate respiratory pathogen spread across large populations.

\subsection{Comparison with Other ABMs of Infectious Disease Spread}

A wide range of ABMs have been developed to simulate infectious disease dynamics, varying in scale, granularity, data inputs, and intervention support. Among these, Corvid \cite{chaoModelingLayeredNonpharmaceutical2020} is the most structurally similar to Epicast. Originally adapted from FluTE \cite{chaoFluTEPubliclyAvailable2010}, which itself was based in part on earlier versions of Epicast \cite{germannMitigationStrategiesPandemic2006}, Corvid uses a discrete community structure and a distributed exposure model. It supports various interventions including vaccination, antivirals, school closures, and isolation, and is capable of simulating pathogen spread across the entire US population. However, it relies on synthetic population and commute data from 2000, limiting its relevance for contemporary policy applications. In contrast, Epicast 2.0 incorporates more recent demographic and mobility data, improving realism for current and future epidemics. Its flexible policy modeling and efficient parallelization also make it better suited to evaluating evolving strategies across a national population.

A different class of ABMs, such as CityCOVID \cite{ozikPopulationDatadrivenWorkflow2021}, EpiSimdemics \cite{barrettEpiSimdemicsEfficientAlgorithm2008}, and EpiHiper \cite{chenEpihiperHighPerformance2025}, explicitly represent agent networks as graphs. CityCOVID models hourly schedules and complex contact structures, supporting detailed non-pharmaceutical interventions (NPIs) and seasonality. However, its high computational cost limits scalability beyond large urban areas like Chicago. EpiSimdemics has simulated populations up to \alttilde 250 million agents \cite{bissetSimulatingSpreadInfectious2012}, but does not appear to be actively maintained. EpiHiper, the apparent successor to EpiSimdemics, has simulated populations up to \alttilde 30 million agents (e.g., California), performs simulations at the state level and aggregates results to approximate national-scale dynamics. This makes modeling long-distance transmission more difficult, as it requires explicit handling of inter-state contacts and travel.

CovidSim \cite{fergusonReport9Impact2020a} is another national-scale ABM used during the COVID-19 pandemic. While it supports key NPIs, such as social distancing, quarantine, and closures, it does not support multi-node parallelism, limiting its scalability on high-performance computing (HPC) infrastructure. Additionally, it lacks pharmaceutical intervention modeling such as vaccination or therapeutics.

Several smaller-scale but feature-rich ABMs, such as Covasim \cite{kerrCovasimAgentbasedModel2021}, OpenABM-Covid19 \cite{hinchOpenABMCovid19AnAgentbasedModel2021}, and OpenCOVID \cite{shattockImpactVaccinationNonpharmaceutical2022}, offer detailed representations of NPIs, testing, contact tracing, and vaccination. These models are often used to study dynamics in localized settings (e.g., cities or regions) but lack the scalability or integration needed to simulate national-level transmission patterns and policy interactions.

In summary, while many ABMs offer strengths in particular domains, realistic contact structures, detailed NPIs, or large population coverage, Epicast 2.0 fills a unique role in combining: up-to-date demographic and mobility inputs, support for both pharmaceutical and non-pharmaceutical interventions, efficient scalability to the full US population, and multi-node paralellism for HPC environments. Table \ref{tab:abm_comparison} summarizes the ABMs described in this section. 

\begin{table}[ht]
\centering
\caption{Comparison of selected agent-based models (ABMs) for infectious disease spread}
\label{tab:abm_comparison}
\small
\resizebox{\textwidth}{!}{%
\begin{tabular}{l p{2.5cm} p{2.7cm} p{2cm} p{2cm} p{5.5cm}}
\toprule
\textbf{Model} & \textbf{Scale} & \textbf{Interventions} & \textbf{Year of Data} & \textbf{Parallelism} & \textbf{Notes} \\
\midrule
\textbf{Epicast 2.0} & National (\alttilde 324M agents) & NPIs, vaccines, \hspace{.18in}antivirals & 2019 & Multi-node HPC & Fully parallelized; supports irregular travel and targeted interventions \\
\textbf{Corvid} & National (\alttilde 280M agents)& NPIs, vaccines, \hspace{.18in}antivirals & 2000 census & Yes & Based on FluTE; community-based exposure model; outdated demographic data \\
\textbf{CityCOVID} & City-level (Chicago) & NPIs & 2020 & Yes & High-resolution agent schedules; graph-based; not optimized for national scale \\
\textbf{EpiHiper} & State-level (up to 30M agents) & NPIs & Varies by study & Yes & Graph-based network model; state-level simulations aggregated for national analysis \\
\textbf{CovidSim} & National (US, UK) & NPIs & 2020 & Single-node only & HPC-capable but lacks distributed parallelism; no vaccine or therapeutic modeling \\
\textbf{Covasim} & Local/regional & NPIs, vaccines, \hspace{.18in}testing, tracing & 2019 & Yes & Rich feature set; less suited for nationwide simulations \\
\textbf{OpenABM-Covid19} & Local/regional & NPIs, vaccines, \hspace{.18in}testing, tracing & 2011 & Yes & Microscale dynamics focus; not intended for full national integration \\
\bottomrule
\end{tabular}
}
\end{table}

\section{Model Overview and New Features}

The agent-based modeling framework that forms the basis of Epicast was originally developed to support bioterrorism preparedness and influenza pandemic planning \cite{halloranContainingBioterroristSmallpox2002, germannMitigationStrategiesPandemic2006}. 
Epicast 2.0 is a large-scale, discrete-space ABM designed to simulate the spread of respiratory pathogens (e.g., influenza and SARS-CoV-2) in the United States. It builds on the original Epicast framework with several major enhancements that improve demographic realism, structural fidelity, and policy flexibility. In this section, we first summarize the key new and updated features introduced in Epicast 2.0, and then describe the model architecture and simulation logic in detail.

\subsection{New and Updated Features in Epicast 2.0}

Epicast 2.0 incorporates several important improvements over earlier versions:

\begin{itemize}
    \item \textbf{Updated synthetic population:} Epicast 2.0 uses a high-fidelity synthetic population of \alttilde 324 million agents generated by the UrbanPop framework \cite{tuccilloUrbanPopSpatialMicrosimulation2023}. Each agent has 25 attributes, including household composition, age, race, ethnicity, school grade, and industry of employment, matched to US Census data as of 2019.
    
    \item \textbf{Enhanced work and school assignment:} A new data-driven algorithm assigns employed agents to workplace networks and students to classrooms (school-groups). Teachers are explicitly modeled and assigned to schools, allowing more realistic simulation of education-sector dynamics.
    
    \item \textbf{Industry-specific exposure scaling:} Contact rates for co-workers and public-facing employees are now scaled using industry-specific survey data, allowing differentiation in occupational transmission risks.
    
    \item \textbf{Flexible, localized policy engine:} The policy model has been updated to support time-varying interventions tailored by geography (e.g., county), industry, or school type. Users can simulate policies such as targeted closures, capacity restrictions, or masking mandates.
\end{itemize}

Together, these enhancements allow Epicast 2.0 to simulate complex and heterogeneous pandemic dynamics with greater resolution and realism.

\subsection{Model Framework} \label{sec:epicast_framework}

Epicast 2.0 operates on a synthetic population that can represent the entire United States or any subset of states. Each census tract is divided into spatial units called \textit{communities}, which serve as the basic units of simulation. Communities are designed to hold approximately 2,000 agents and function as the primary locations for agent interaction and pathogen transmission.

Agents are assigned to a \textit{nighttime} (residential) community and, for working adults, a \textit{daytime} (workplace) community. Simulations proceed in two 12-hour time steps per day. During each step, agents interact with others within their current community and associated social venues (e.g., households, schools, workplaces). After each time step, agents move to their alternate location (from home to work/school or vice versa), and the interaction and transition process repeats.

Pathogen transmission is determined probabilistically based on contact structure, venue-specific exposure intensity, and pathogen properties (e.g., infectiousness). Epicast 2.0 also models long-distance travel (e.g., vacations or infrequent trips), which introduces opportunities for seeding outbreaks across geographically distant communities. In addition, a flexible policy framework allows users to model detailed interventions that modify agent movement or interaction behavior at various levels of granularity.

To achieve scalability for full US simulations, Epicast makes two key modeling simplifications. First, inter-community interactions cannot occur, agents only interact with agents in their current community and must travel to another community to mix with others there. Second, interactions within venues assume all-to-all mixing, which reduces the memory and computational burden associated with explicitly modeling contact networks. These design choices, combined with support for HPC environments, allow Epicast to run efficiently at national scale. This flexible and computationally efficient framework has enabled Epicast to support both retrospective and prospective analyses of pandemic scenarios \cite{delvalleEpiCastSimulatingEpidemics2021, germannAssessingK12School2022}.

\begin{figure}[h]
    \centering
    \includegraphics[width=16.5cm]{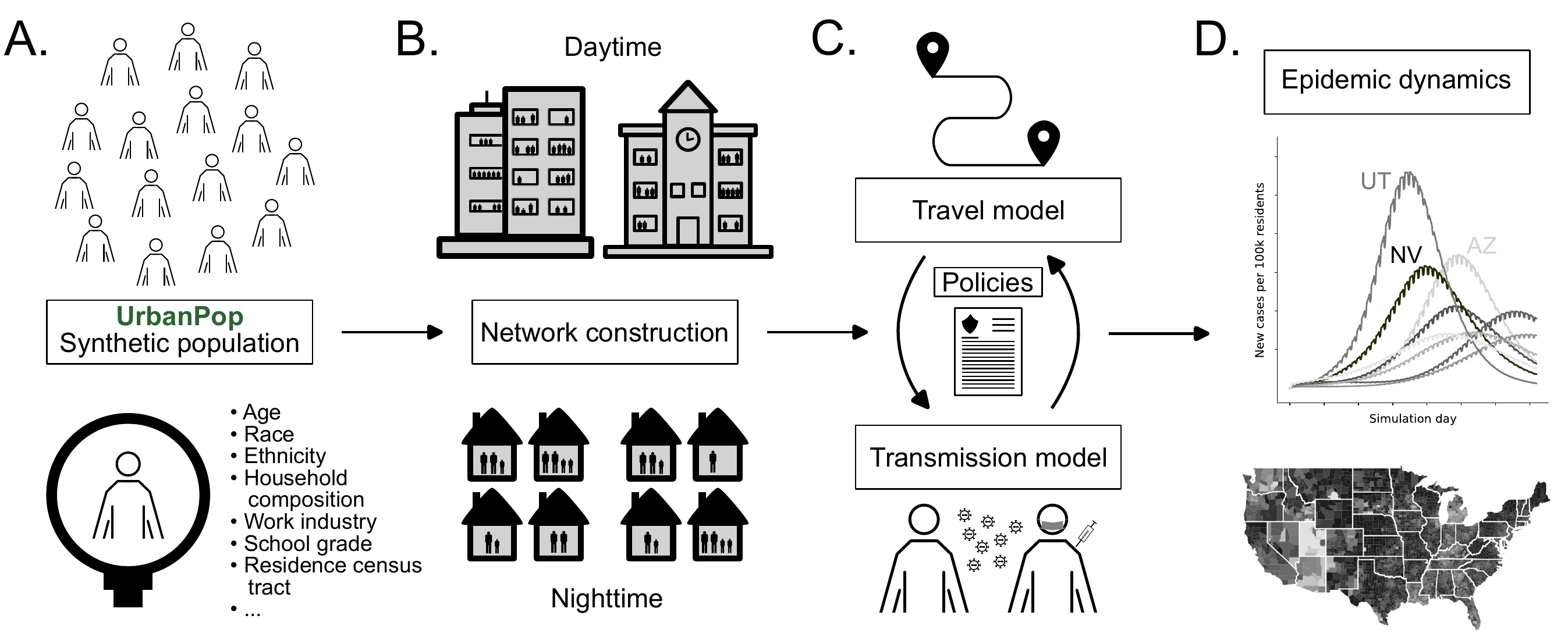}
    \caption{Schematic overview of Epicast 2.0. \textbf{A}) UrbanPop provides a synthetic population of \alttilde 324 million agents with detailed demographic information. \textbf{B}) Epicast 2.0 constructs nighttime and daytime networks in which agents interact with other agents. \textbf{C}) Travel models specify how agents are reorganized from their nighttime to daytime networks as well as the details of longer distance trips. User specified policies can modify agent's movement (by e.g. closing schools) and encourage or enforce non-pharmaceutical interventions. A transmission model specifies how pathogens spread between agents. \textbf{D}) Spatio-temporal dynamics of the resulting epidemic can be analyzed down to the level of individual agents and their demographics.}
    \label{fig:schematic}
\end{figure}

Figure~\ref{fig:schematic} provides a schematic overview of the Epicast 2.0 architecture.

\subsection{Synthetic Population} \label{synthetic-population}

Epicast 2.0 uses a synthetic population of approximately 324 million agents, covering all 50 US states and the District of Columbia. The population is generated using UrbanPop \cite{tuccilloUrbanPopSpatialMicrosimulation2023}, a spatial microsimulation framework that combines Public Use Microdata Sample (PUMS) records with American Community Survey (ACS) data to produce demographically accurate agent-level data.

Each agent in the population is assigned 25 attributes (listed in Table \ref{tab:agent_attrs}), including age, sex, race, ethnicity, household composition, income, employment status, industry of employment, school grade, and commuting behavior. These attributes are critical for capturing individual-level heterogeneity in disease risk, contact patterns, and policy responsiveness.

\begin{table}[h]
    \begin{center}
    	\begin{tabular}{@{}lcl@{}} 
    		\toprule
    		\thead{Attribute} & \thead{Data type / Range} & \thead{Notes} \\
    		\midrule
            Geographic ID & Numeric identifier & FIPS code of agent's residential block-group \\ 
    		Person ID & Numeric identifier & Globally unique agent identifier \\ 
    		Household ID & Numeric identifier & Globally unique household identifier \\ 
    		Household income & 0 - 999,999,999 & Total household income per year \\ 
    		Household size & 1 - 20 &  Total number of agents living in household \\ 
    		Household type & Categorical & \makecell{Group housing, single- or multi-family, \\ mobile home, etc.} \\ 
    		Household kids & Boolean & Household has child residents \\ 
    		Household workers & 0 - 20 & Total number of employed agents in household \\ 
    		Household non-workers & 0 - 20 & Total number of non-workers in household \\ 
    		Household adult workers & 0 - 20 & Number of adult employed agents in household \\ 
    		Household adult non-workers & 0 - 20 & Number of adult non-workers in household \\ 
    		Household arrangement & Categorical & Married, no spouse, alone, not alone \\
    		Household tenure & Categorical & Rent, own, other \\
    		Household vehicles & Ordinal & Number of vehicles household has access to, 0 - 6+ \\
    		Agent age & 0 - 99 & Agent age in years \\
    		Agent sex & Categorical & Male or female \\
    		Agent race & Categorical & \makecell{Black, White, Asian, Native American, \\ Pacific Islander, other, multi} \\
    		Agent ethnicity & Categorical & Hispanic or non-Hispanic \\
    		Agent IPR & Ordinal & Income-poverty-ratio, 6 levels between 0.0 and >2.0 \\
    		Agent employment & Categorical & Employed, unemployed, military, not in labor force \\
    		Industry of employment & Numeric identifier & 3 digit NAICS code of employment industry \\ 
    		Commute time & 0 - 200 & Commute time in minutes \\
    		Commute mode & Categorical & \makecell{NA, private vehicle, public transportation, \\ bicycle, walk, etc.}  \\ 
    		Commute occupancy & Categorical & NA, drove alone, carpool \\ 
    		Agent school grade & Ordinal & Preschool through graduate school \\
    	\end{tabular}
        \caption{UrbanPop synthetic agent attributes.}
        \label{tab:agent_attrs}
    \end{center}
\end{table}

UrbanPop works by upsampling PUMS records, which capture the joint distribution of demographic variables at the Public Use Microdata Area (PUMA) level, so that the resulting synthetic population matches the marginal distributions of these variables at the census block-group level, as provided by the ACS. This ensures that both joint and marginal distributions of demographic variables in the synthetic population reflect those observed in real US Census data.

As a result, Epicast 2.0’s synthetic population preserves important real-world correlations (e.g., between age and household size, or between race, income, and occupation) that influence both disease spread and the impact of interventions. For this study, the synthetic population reflects US demographics as of 2019.

\subsection{Community Assignment and Network Construction} \label{network-construction}

Epicast 2.0 divides each census tract into smaller geographic units called communities, each containing approximately 2,000 agents. Communities serve as the fundamental spatial unit for simulation, agents live, work, attend school, and interact within communities, and transmission occurs only between agents located in the same community during a given time step. The network of agents has two states:
\begin{itemize}
    \item \textbf{A nighttime (residential) community}, determined by the agent's household location in the synthetic population.
    \item \textbf{A daytime community}, for employed adults, determined by a commute model (described in \nameref{commute-model}).
\end{itemize}

If a census tract contains more agents than can fit into a single community, multiple communities are created. Additionally, “work-only” communities are added to represent areas (e.g., city centers or industrial zones) that receive a large influx of commuters (more than \alttilde 1,000) but have few or no residents.

Agents are grouped within communities based on their residential block-group and household membership to preserve geographic and social clustering. Only working adults travel to different communities during the day; children, older adults, and unemployed individuals remain in their residential community throughout the day.

Within each community, agents are organized into \textit{venues}, which represent distinct social or spatial interaction contexts. Venues are hierarchical and include: Households, Household clusters, Schools and classrooms (school-groups), Workplaces and work-groups, Neighborhoods, and General community spaces. Note that the community is the largest venue, and each community contains four neighborhoods (the next largest venue) into which households are assigned in approximately equal number. In general, smaller and more specific venues (e.g., households or work-groups) are assumed to have higher-intensity interactions, while broader venues (e.g., neighborhoods or general community) have lower-intensity, more diffuse contact. These interaction intensities are user-configurable.

\subsubsection{Schools and School-groups} \label{school-groups}

Each community in Epicast 2.0 includes a set of simulated educational institutions that serve as both structural and epidemiological venues for student interaction. By default, each community contains:

\begin{itemize}
    \item One \textbf{high school} (grades 9–12)
    \item One \textbf{middle school} (grades 6–8)
    \item Two \textbf{elementary schools} (grades 1–5, each serving two neighborhoods)
    \item One \textbf{kindergarten}
    \item One \textbf{preschool}
\end{itemize}

Student agents are assigned to schools based on their school grade, as specified in the synthetic population. All school-age children attend school within their residential community. (Note: this assumption is due to current limitations in the \nameref{commute-model} and may be relaxed in future iterations.)

To better capture classroom-level mixing, students within each school are further grouped into \emph{school-groups}, which represent classrooms or cohorts where interaction intensity is higher than in the general school environment. The number of school-groups in a school is determined based on county-level student–teacher ratios for that school type derived for the entire US (see \nameref{school-group-data}). Once the number of school-groups is determined, students are randomly assigned to each group.

\textbf{Teacher agents}, identified by employment in the educational services sector (NAICS code 611), are assigned to daytime communities by the \nameref{commute-model} and are then randomly allocated to specific schools, and then to school-groups. Each school-group is assigned one teacher, where possible. If there are more teachers than school-groups in a school, excess teachers are grouped into a single school-group corresponding to an administrative unit, representing non-classroom personnel (e.g., counselors, principals, or aides) who interact with one another at relatively high intensity.

\subsubsection{Workplaces and Work-Groups} \label{work-group-assign}

Adult agents employed in non-education sectors are assigned to workplaces and work-groups based on their industry (using NAICS codes) and their daytime community, as determined by the \nameref{commute-model}. Each industry in each community is subdivided into work-groups, which represent direct co-worker teams -- analogous to departments or project teams. The number of work-groups is determined using a database of industry-specific work-group sizes (see \nameref{work-group-data}), derived from national labor survey data. Workers within each industry are randomly assigned to work-groups of the appropriate size.

Work-groups are then geographically allocated to one of the community’s neighborhoods. All work-groups of a given industry within the same neighborhood are bundled into a workplace, which approximates a shared business facility or office complex. This two-tiered structure captures both high-intensity intra-group interactions and more diffuse inter-group contact within shared workspaces.
This organizational structure allows Epicast 2.0 to model differences in industry contact structure, supporting more precise estimation of risk and policy impact across industries.

\subsubsection{Household-clusters}

Within each community, households are grouped into \emph{household-clusters} which include small, high-contact social units consisting of four households each. These clusters do not necessarily align with neighborhoods or geographic proximity. Instead, they are intended to approximate informal but epidemiologically relevant contact networks, such as extended families, close friend groups, or shared childcare arrangements. These contacts are assumed to persist even under most restrictive policies, and thus serve as a mechanism for continued transmission when control policies are in effect (e.g., when schools are closed).

\subsection{Travel Models} \label{travel-models}
Epicast 2.0 incorporates two models that govern the movement of agents across communities:
\begin{enumerate}
    \item \textbf{Regular travel mode}: Simulates daily commuting patterns, typically weekdays, in which employed adults travel between their nighttime (residential) communities and daytime (work) communities.
    \item \textbf{Irregular travel mode}: Captures infrequent, longer-distance trips for business or leisure purposes such as business, vacation, or family visits. These trips span regional and national distances and may occur via road, air, or rail.
\end{enumerate}

Together, these two models capture both short-range, high-frequency mobility and long-range, low-frequency mixing, both of which are critical for understanding spatial transmission dynamics.

\subsubsection{Regular travel model} \label{commute-model}
The regular travel model in EpiCast 2.0 assigns daytime (work) communities to all employed, adult agents including teachers. Although teachers, like students, participate in school-groups rather than work-groups, unlike students, teachers can commute to schools outside of their nighttime community. 

At the core of the regular travel model is tract-to-tract flow matrix, a sparse connectivity matrix that represents \emph{worker flow} derived from the Longitudinal Employer-Household Dynamics (LEHD) Origin-Destination Employment Statistics (LODES) version 7 \cite{lodes2019}.  While LODES7 reports flows between 2010 census blocks (not conditioned on any demographics), Epicast requires alignment with 2019 tract-level geography to match the UrbanPop synthetic population, which is based on 2019 American Community Survey (ACS) data. Therefore, worker flows must be mapped from 2010 block-level FIPS codes to 2019 tract-level FIPS codes. As the set of 2019 geographic units (tract, blocks, etc.) are intermediate between the 2010 and 2020 systems, this mapping requires a "crosswalk" that specifies the weighted contribution of the 2010 census blocks to the 2020 census tracts. This crosswalk, which we refer to as $xwalk$ below, is provided by the IPUMS NHGIS \cite{mansonNationalHistoricalGeographic2024}.

\begin{figure}[h]
    \centering
    \includegraphics[width=6cm]{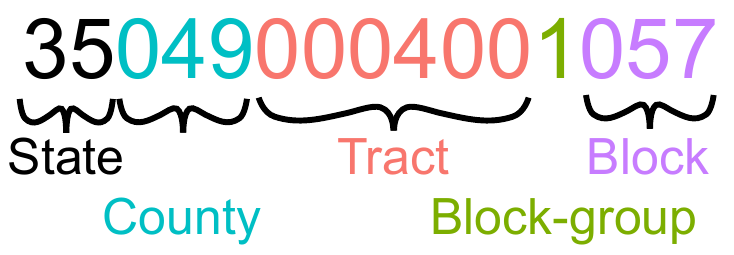}
    \caption{Structure of a 15-digit, block-level Federal Information Processing Standard (FIPS) code. Tract-level FIPS codes are 11 digits and are extracted by removing the final 4 digits.}
    \label{fig:fips}
\end{figure}

The conversion map $LODES7_{block} \rightarrow UrbanPop_{tract}$ was constructed by the following steps:

\begin{enumerate}
	\item \textbf{Convert blocks to tracts:} Extract all origin ($src$) and destination ($dest$) tracts from the LODES7 dataset by converting block-level FIPS codes to 2010 tract-level FIPS codes (by removing the last four digits as shown in Figure \ref{fig:fips})).
	\item \textbf{Direct match: }If an UrbanPop tract appears among the converted $src$ or $dest$ tracts, then no further mapping is required. This means that tract's geographic definition did not change between 2010 and 2019.
	\item \textbf{Crosswal-based mapping:} If an UrbanPop tract is not directly matched, but it is present within the destination 2020 census tracts of $xwalk$, then apply the $xwalk$ 2010 block $\rightarrow$ 2020 tract mapping, using the weights provided in $xwalk$ to reallocate worker flows accordingly.
	\item \textbf{Parent-child aggregation fallback:} If the UrbanPop tract is not present in $xwalk$, but \emph{child tracts} (i.e., related 2019 tracts with the same high-order FIPS digits) are present, then use the children's mappings. Sum the worker outflows and inflows to construct the flow data for the parent tract.
	\begin{itemize}
		\item \textbf{A parent tract:} Two tracts have a \emph{parent-child} relationship if they share at least the high four digits of the six tract digits of a FIPS code with one or more children, and its remaining digits are zero.
		\item For example, tract 36053030600 (in the 2019 ACS) didn't exist in 2010 or 2020. However, its child tracts [36053030601, 36053030602] do exist in 2020. Therefore, all 2010 blocks that map to either child (36053030601 or 36053030602) are reassigned to the parent UrbanPop tract 36053030600
	\end{itemize}
\end{enumerate}

This process ensures that all worker flows are translated into the geographic framework used by UrbanPop, preserving realistic commuting patterns across space and census definitions.

Once the \( LODES7_{block} \rightarrow UrbanPop_{tract} \) map is constructed, tract-level worker flows are computed by aggregating LODES7 block-level flows according to their mapped destination and origin tracts. This process successfully assigns inflow and outflow values to nearly all tracts in the UrbanPop population. The only exception is a set of three tracts in Oglala Lakota County, South Dakota, which could not be mapped due to missing or ambiguous geographic identifiers. For these, we imputed worker flow values by copying those of the nearest tract within the same county.

\begin{figure}[htb]
    \centering
    \includegraphics[width=16cm]{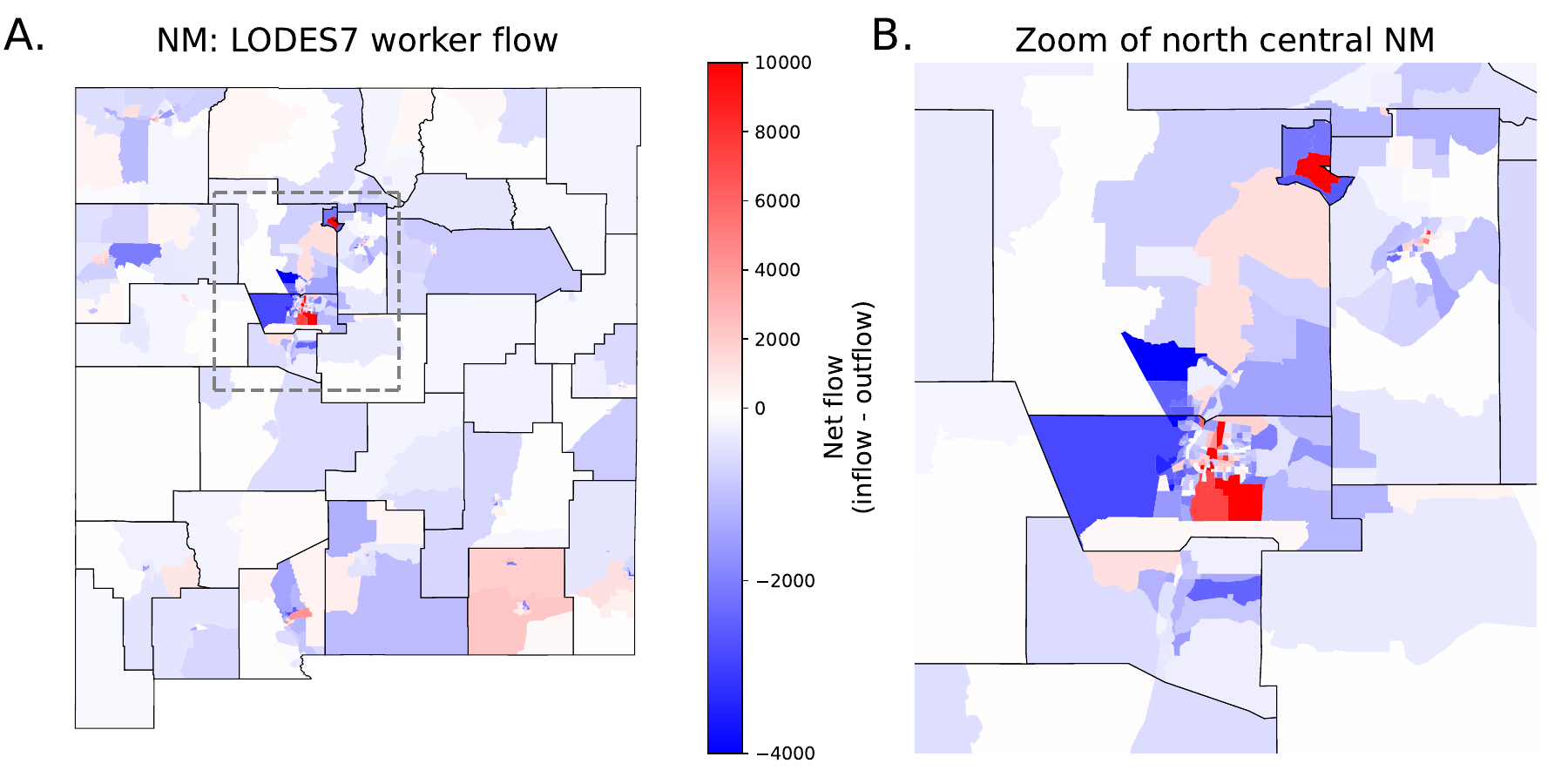}
    \caption{Worker flow patterns in New Mexico. \textbf{A)} Choropleth heatmap of net worker flow for each census tract in New Mexico (inflow minus outflow). Note that the color map is not symmetric. County boundaries are shown in black. \textbf{B)} Zoom of \textbf{A} in the area of Los Alamos, Santa Fe, Bernalillo, Valencia, and Sandoval counties highlighting the flow of workers between more residential tract (blue) and tracts with more work places (red). For reference, the strongly red tracts in the center of the image are located in and around the city of Albuquerque, the largest city in New Mexico. The strongly red tract in the center-top of the image contains Los Alamos National Lab, a major employer in the region. }
    \label{fig:workerflow}
\end{figure}

To further improve realism, we restrict interstate commuting to occur only between states that share a border or are within "reasonable" commute distance from each other (e.g. Connecticut and New Jersey). This constraint eliminates a small number of implausible long-distance interstate flows that appear in the raw LODES data but are unlikely to represent routine daily commutes. Figure \ref{fig:workerflow} illustrates the results of this process for the state of New Mexico.

Within Epicast 2.0, the worker flow connectivity matrix is used to assign \textbf{work (daytime) communities} to all employed, adult agents living within a given tract $T$ using the following procedure:

\begin{enumerate}
    \item For the agent's home tract $T$, extract the vector of outbound worker flows to all other tracts
    \item Convert the outflow vector to a cumulative sum and normalize by total outflow to generate an empirical cumulative distribution function (eCDF) 
    \item For each agent:
    \begin{enumerate}
        \item Assign a work tract by inverse transform sampling of the eCDF.  
        \item Assign a work community by uniform random sampling from the communities that belong to the chosen tract.
    \end{enumerate}
\end{enumerate}

\subsubsection{Irregular travel model} \label{ldtm}

Epicast 2.0 includes an irregular travel model that simulates infrequent, long-distance trips such as vacations, business travel, or family visits. These trips contribute to long-range spatial mixing and outbreak seeding and are governed by three components: 
\begin{enumerate}
    \item Timing: when a trip occurs 
    \item Destination: where the agent travels 
    \item Duration: how long the agent remains away
\end{enumerate}

The probability that an agent initiates a trip on any given time step is given by $P_\text{travel} \cdot P_\text{travel}(age)$, where $P_\text{travel}$ is a user-defined parameter uniformly adjusts a set of built-in, age-specific baseline probabilities $P_\text{travel}(age)$ that are listed in Table~\ref{tab:travel_probs}. Agents who are currently traveling or are otherwise ineligible (e.g., hospitalized or quarantined) are excluded from irregular travel.

\begin{figure}[h]
    \centering
    \includegraphics[width=16cm]{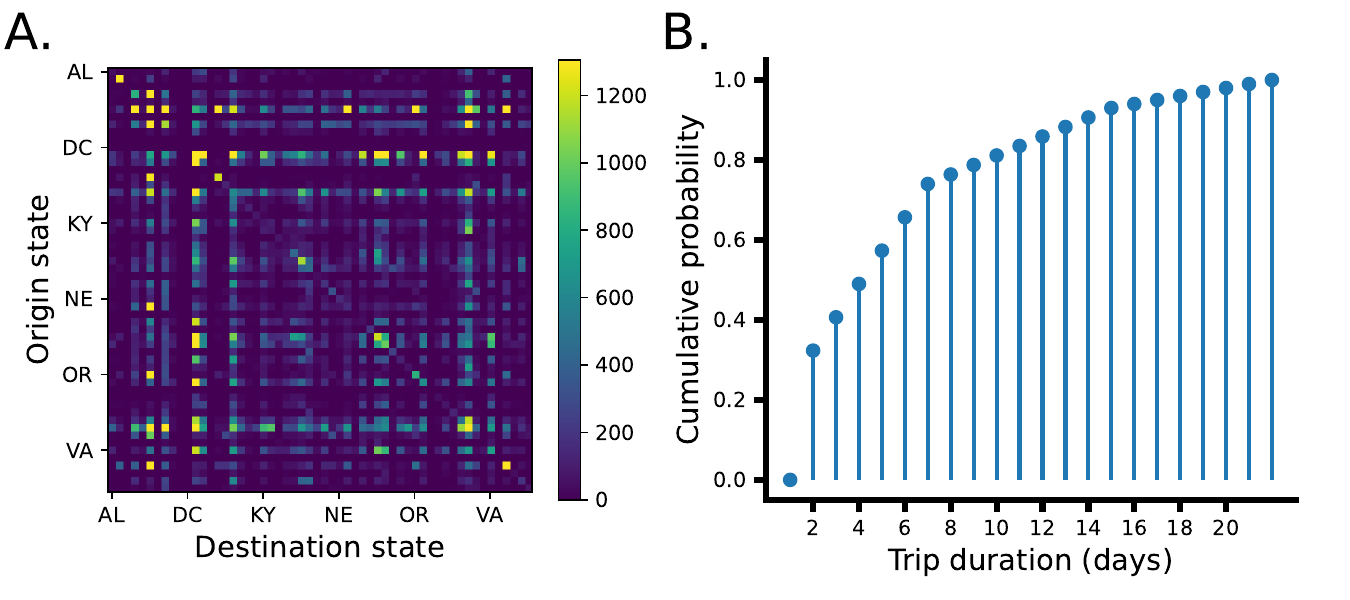}
    \caption{Primary data underlying the irregular travel model. \textbf{A)} State-to-state travel connectivity matrix based on the number of flights within and between each state and Washington DC in 2019. \textbf{B)} Cumulative distribution of irregular trip durations. }
    \label{fig:travel}
\end{figure}

When an agent embarks on a trip, their destination is determined by a connectivity matrix comprised of flight count data for airline flights within and between each US state (and Washington DC) during 2019 \cite{bureauoftransportationstatisticsTranStats2019}. This irregular travel matrix is shown in Figure \ref{fig:travel}A. The vector of outflows from the traveling agent's home state is converted to an eCDF and a destination state is selected by inverse transform sampling, analogous to the procedure used within the regular travel model. The agent's destination tract and community are then determined by uniform random sampling of tracts within the destination state and communities within the selected tract. The duration of the trip is then determined by inverse transform sampling of an eCDF of trip durations shown in Figure \ref{fig:travel}B. During travel, the agent resides in the destination community and continues participating in the standard day/night venue interactions. Once the trip ends, they return to their home community and resume normal activity.

\begin{table}[h]
    \begin{center}
    	\begin{tabular}{@{}ccccc@{}}
    	\toprule
    	\thead{Age 0-5} & \thead{Age 6-17} & \thead{Age 8-29} & \thead{Age 30-64} & \thead{Age 65+} \\
    	\midrule
    	0.0023 & 0.0023 & 0.0050 & 0.0053 & 0.0028 \\
    	\end{tabular}
     \caption{Age-group specific baseline travel probabilities.}
     \label{tab:travel_probs}
    \end{center}
\end{table}

\subsection{Transmission and disease model} \label{transmission-model}

\begin{figure}[h]
    \centering
    \includegraphics[width=16cm]{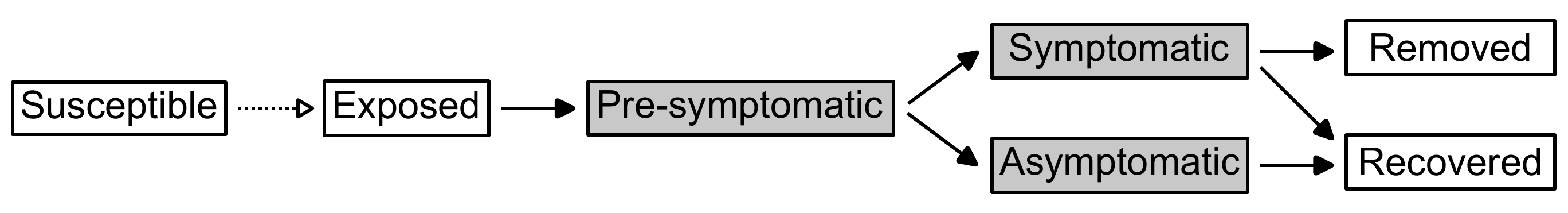}
    \caption{Compartmental structure of the transmission and disease model in Epicast 2.0. Boxes represent health states that agents may occupy during a simulation. Gray-shaded compartments indicate infectious states where agents can transmit the pathogen (unless shedding is disabled in that compartment via user settings). Solid arrows represent transitions that occur deterministically (i.e., with probability 1.0 or summing to 1.0 over time), while dashed arrows denote stochastic transitions that depend on an agent’s exposure level. For clarity, sub-compartments such as hospitalization, isolation, and quarantine are omitted from the figure but may exist within the symptomatic and/or asymptomatic branches. See text for additional configuration details and transition rules.}
    \label{fig:compartments}
\end{figure}

Epicast 2.0 uses a flexible, compartmental transmission model to simulate the progression of infection and the effect of interventions. The model includes both core infectious states and optional sub-compartments (e.g., hospitalization, isolation), which allow it to represent a range of respiratory pathogens. The main compartment structure is shown in Figure \ref{fig:compartments}.

\subsubsection{Susceptible}

At initialization all agents begin in the Susceptible compartment. Agents seeded as index cases are moved directly into the Exposed state, while agents designated as initially immune enter the Recovered compartment (see Section \nameref{index-cases}).

Susceptible agents follow their default daily schedule -- commuting between residential and work/school communities on weekdays and remaining local on weekends -- unless altered by user-defined policies (see Section \nameref{policies}). During each time step, susceptible agents may accumulate exposure based on contact with infectious individuals in the same venues. The amount of accumulated exposure determines the probability of transitioning to the Exposed state; transmission is stochastic and not guaranteed.

\subsubsection{Exposed}

\begin{figure}[h]
    \centering
    \includegraphics[width=13cm]{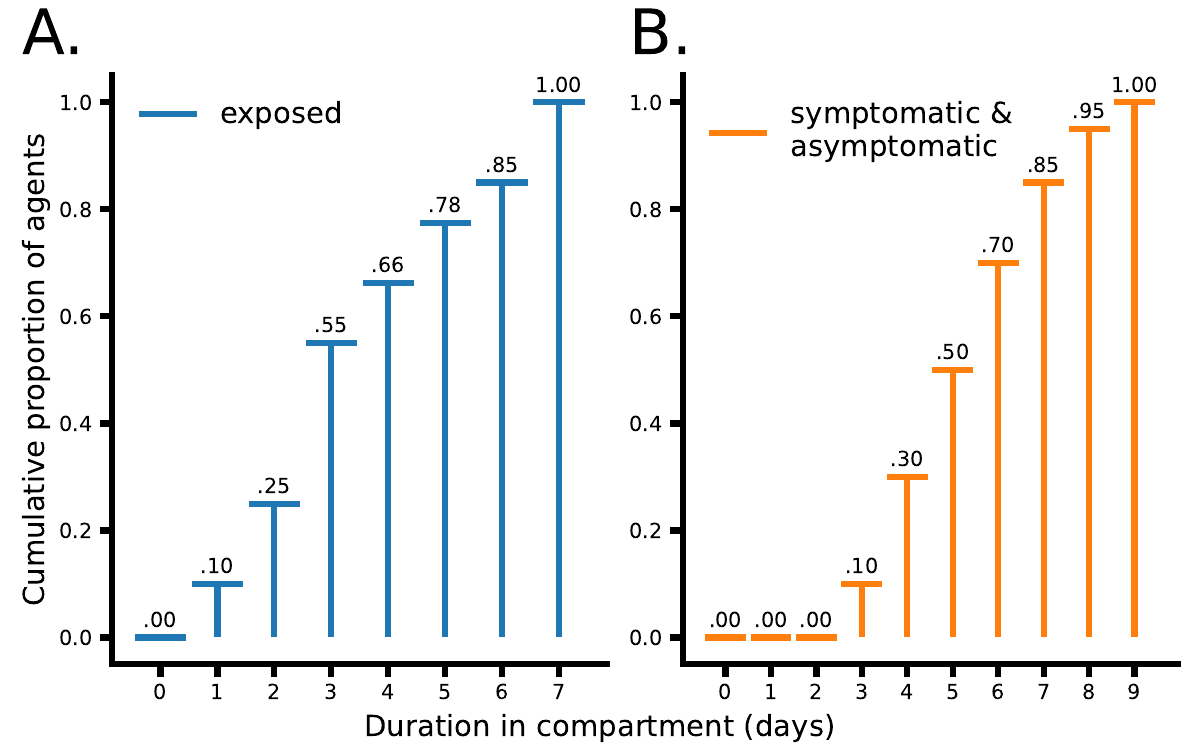}
    \caption{Probability of transitioning out of Exposed (panel \textbf{A}, blue) and Symptomatic/Asymptomatic (panel \textbf{B}, orange) compartments given how long a agent has been in that state (i.e. has not yet transitioned out).}
    \label{fig:compartment_durations}
\end{figure}

Agents in the Exposed state are infected but not yet infectious (i.e., not shedding). This state lasts between 1 and 7 days. On each day, agents have a defined probability of transitioning to the Pre-symptomatic compartment, as shown in Figure \ref{fig:compartment_durations}A.

\subsubsection{Pre-symptomatic}

All agents pass through the Pre-symptomatic compartment, which represents a single day of non-symptomatic shedding. This accounts for pathogens like SARS-CoV-2 that exhibit pre-symptomatic transmission. Shedding intensity in this state is user-configurable and defined relative to symptomatic shedding. Agents then transition to either the Asymptomatic or Symptomatic compartment. The probability of becoming asymptomatic is also user-defined. Setting this probability to zero eliminates the asymptomatic branch.

\subsubsection{Asymptomatic}

The Asymptomatic compartment represents non-symptomatic, infectious individuals. Duration is variable (3–9 days, see Figure \ref{fig:compartment_durations}B), and agents transition directly to Recovered. Shedding intensity is user-defined and shared with the Pre-symptomatic compartment. Note that, in the current implementation, a single non-symptomatic shedding parameter governs the degree of shedding in both the Pre-symptomatic and Asymptomatic compartments (as both are non-symptomatic states). All agents in this state are assumed to remain in circulation (i.e., not withdraw).

\subsubsection{Symptomatic}

The Symptomatic compartment represents agents who exhibit symptoms and are infectious. Duration ranges from 3 to 9 days (same as Asymptomatic, see Figure \ref{fig:compartment_durations}B). Shedding intensity is user-configurable and serves as the baseline for other states.
On the second day of symptoms, agents may transition into healthcare sub-compartments (Hospitalized, ICU, or Ventilated), based on age-specific probabilities (Table \ref{table:table3}, column one). These transitions occur only once, and only on that day. Once hospitalized, agents can transition to the ICU or Ventilated sub-compartments based on the probabilities given in Table \ref{table:table3}, columns two and three. Hospitalized agents interact only with household members and are considered withdrawn from community mixing (see Section \nameref{withdrawn}). 

\begin{table}[h]
    \begin{center}
    	\begin{tabular}{@{}lcccccc@{}}
    	\toprule
        \thead{ } & \thead{Symptomatic\\$\rightarrow$\\Hospitalized} & \thead{Hospitalized\\$\rightarrow$\\ICU} & \thead{ICU\\$\rightarrow$\\Ventilated} & \thead{Hospitalized\\$\rightarrow$\\Removed} & \thead{ICU\\$\rightarrow$\\Removed} & \thead{Ventilated\\$\rightarrow$\\Removed}\\
    	\midrule
    	\textbf{Age 0-5} & 0.0104 & 0.24 & 0.12 & 0.0 & 0.0 & 0.2 \\
        \textbf{Age 6-17} & 0.0104 & 0.24 & 0.12 & 0.0 & 0.0 & 0.2 \\
        \textbf{Age 18-29} & 0.07 & 0.24 & 0.12 & 0.0 & 0.0 & 0.2 \\
        \textbf{Age 30-64} & 0.28 & 0.36 & 0.22 & 0.0 & 0.0 & 0.45 \\
        \textbf{Age 65+} & 1.0 & 0.35 & 0.22 & 0.0 & 0.26 & 1.0 \\

    	\end{tabular}
     \caption{Age-group specific transition probabilities for Symptomatic sub-compartments.}
     \label{table:table3}
    \end{center}
\end{table}

\subsubsection{Withdrawn} \label{withdrawn}

Agents can enter Withdrawn state in two ways: \emph{voluntarily}, if an agent is symptomatic (with age-specific probabilities over the first 3 days; Table \ref{table:table4}), or \emph{obligatorily} if an agent is hospitalized (including ICU or ventilated sub-states). Withdrawn agents interact only with members of their household and remain withdrawn until they recover or are removed.

\begin{table}[h]
    \begin{center}
    	\begin{tabular}{@{}lccc@{}}
    	\toprule
        \thead{ } & \thead{Day 0} & \thead{Day 1} & \thead{Day 2} \\
    	\midrule
    	\textbf{Age 0-5} & 0.3 & 0.8 & 0.9 \\
        \textbf{Age 6-17} & 0.3 & 0.7 & 0.8 \\
        \textbf{Age 18+} & 0.3 & 0.5 & 0.7 \\ 
    	%\bottomrule
    	\end{tabular}
     \caption{Age-group specific withdrawal probabilities.}
     \label{table:table4}
    \end{center}
\end{table}

\subsubsection{Recovered}

Agents recover from either the Asymptomatic or Symptomatic compartment (unless removed via mortality). Recovered agents may return to Susceptible if reinfection is allowed (partial or full, based on user settings). Otherwise, they are excluded from further transmission, though not marked as deceased. Note that recovered agents may be susceptible to new variants or all variants (including the variant they were previously infected with) depending on user specification.

\subsubsection{Removed}

Agents who die from the disease enter the Removed compartment and are no longer active in the simulation. If reinfection is disabled, Recovered and Removed are identical except for bookkeeping purposes. Table \ref{table:table3}, columns 4-6 gives the age-specific probabilities that agents transitioning out of the hospitalized sub-compartments will be removed.

\subsubsection{Transmission model}

Agent interactions, which form the basis of Epicast 2.0's transmission model, only occur within communities; there are no inter-community interactions within a time step. 

Epicast 2.0 simulates pathogen transmission through agent interactions that occur within shared communities and shared venues during each 12-hour time step; there are no inter-community interactions. The probability of infection for a susceptible agent depends on the presence of infectious individuals in the same venues, the structure of social mixing, and the pathogen’s transmissibility.

\textbf{Venue Structure}
As described in \nameref{network-construction}, within a community agents can interact within eight different venues: general community, general neighborhood, schools, school-groups, workplaces, work-groups, household-clusters, and households. These venues are not mutually exclusive, agents typically participate in several concurrently (e.g., household, neighborhood, and community). The exceptions are school-related and work-related venues, which are mutually exclusive by design (i.e., an agent is either a student or a worker). For example, if two agents share a household, they will necessarily interact within the household, household-cluster, neighborhood, and community venues (and possibly school or work, but that is relatively rare for members of the same household). 

\textbf{Interaction Intensity Scale Factors (IISF)} 
Interactions within a venue are assumed to be all-to-all, meaning every agent in the venue is considered to potentially interact with every other agent present. As a result, the intensities of those interactions are crucial. Within Epicast 2.0, interaction intensity approximates the contribution of the duration and proximity of an interaction to the transmission process (i.e., a pathogen independent contribution), and is specific to venues and age-groups.

Each venue has an associated interaction IISF, which reflects the average contact intensity within that space, accounting for factors like proximity and duration. IISF values vary by:

\begin{itemize}
    \item Venue type (e.g., household vs. community)
    \item Age pairings (e.g., child–child vs. adult–child)
    \item Geographic location (if customized by the user)
\end{itemize}

In general, IISF values are higher in smaller, more intimate venues (e.g., households), and lower in larger, more diffuse venues (e.g., communities). Tables \ref{table:iis_age} and \ref{table:iis_school} provide the baseline IISF values used in Epicast 2.0. for the venues in which all age-groups mix (household, household-cluster, neighborhood, and community) and for schools, respectively. The work venue has a single baseline IISF, 0.07475, as all working agents are assumed to be adults.

\begin{table}[h]
    \begin{center}
    	\begin{tabular}{@{}lccccc@{}}
            \toprule
            \thead{} & \thead{Age 0-5} & \thead{Age 6-17} & \thead{Age 18-29} & \thead{Age 30-64} & \thead{Age 65+} \\
            \midrule
            \multicolumn{6}{c}{\textit{Household}} \\
            Child & 0.6 & 0.6 & 0.3 & 0.3 & 0.3 \\      
            Adult & 0.3 & 0.3 & 0.4 & 0.4 & 0.4 \\
            \multicolumn{6}{c}{\textit{Household-cluster}} \\
            Child & 0.2475 & 0.2475 & 0.132 & 0.132 & 0.132 \\     
            Adult & 0.132 & 0.132 & 0.165 & 0.165 & 0.165 \\
            \multicolumn{6}{c}{\textit{Neighborhood}} \\
            All & 0.000082 & 0.000247 & 0.00066 & 0.00066 & 0.00099 \\
            \multicolumn{6}{c}{\textit{Community}} \\
            All & 0.000021 & 0.000062 & 0.000165 & 0.000165 & 0.000247 \\
    	\end{tabular}
        \caption{Baseline interaction intensity scale factors (IISF) for venues with full age-based contact patterns. Each row represents a type of shedding agent (adult (18+), child (0-17), or all) within a specific venue, and each column specifies the age-group of the susceptible agent being exposed.}
        \label{table:iis_age}
    \end{center}
\end{table}

\begin{table}[h]
    \begin{center}
    	\begin{tabular}{@{}lcccccc@{}}
            \toprule
            \thead{} & \thead{High school} & \thead{Middle school} & \thead{Elementary\\school} & \thead{Kindergarten} & \thead{Pre-k} \\
            \midrule
            Student $\rightarrow$ student & 0.441 & 0.525 & 0.609 & 0.35 & 0.15 \\      
            Student $\rightarrow$ teacher & 0.2 & 0.25 & 0.3 & 0.3 & 0.3 \\
            Teacher $\rightarrow$ student & 0.2 & 0.25 & 0.3 & 0.3 & 0.3 \\
            %\bottomrule
    	\end{tabular}
        \caption{Baseline IISF for schools.}
        \label{table:iis_school}
    \end{center}
\end{table}

Note that the IISF listed in Tables \ref{table:iis_age} and \ref{table:iis_school} are the baseline IISF within Epicast 2.0 and can be further modified by user-defined, global scaling (per venue). Additionally, these IISF can also be scaled in a spatiotemporally specific manner (e.g., reducing school interactions in one county during a specified time period) by means of user-defined policies (see \nameref{policies}).

\textbf{Exposure Accumulation and Transmission}
At each time step, susceptible agents accumulate exposure from infectious agents within their shared venues, the intensities of which are given by the applicable IISF. The total exposure determines the probability of transitioning to the Exposed state. If a susceptible agents does not transition to Exposed, their accumulated exposure is reset to zero for the start of the next time step.

The probability is computed using:

\begin{equation}
    P_{k} = 1 - \prod_{v}^{venues}{\ \prod_{j \in v}^{agents}{1 - P_\text{trans}X_{j}}} \quad \mathrm{where}
    \label{eq:transition}
\end{equation}

where:

\begin{equation}
    X_{j} = I_{j}\ C_{v}(age_{k},age_{j})\ M_{v,community} \ (1 - VE_{j}) \ldots
    \label{eq:exposure}
\end{equation}

Definitions:
\begin{itemize}
    \item $P_{k}$: Probability that susceptible agent $k$ becomes exposed at the end of the current time step. 
    \item $P_\text{trans} \in \left(0,1\right)$: Global transmissibility constant (pathogen-specific). 
    \item$X_{j}$: Exposure that agent $k$ experiences due to agent $j$ while interacting within venue $v$. 
    \item $I_{j} \in \left[0,1\right]$: Shedding level of agent $j$ (typically 1.0 if $j$ is in the Symptomatic state, lower if $j$ is in the Pre-symptomatic or Asymptomatic states, and zero otherwise).
    \item $C_{v}(age_{k},age_{j})$: Age-specific IISF for venue $v$, between age groups $age_{k}$ and $a_j$. Note that for some venues $C_{v}$ is uniform across age (e.g. work venues), but we include age generally in Equation \ref{eq:exposure} for simplicity. 
    \item $M_{v,community}$: Community-specific IISF for venue $v$ (i.e., for the community that $k$ and $j$ currently occupy).
    \item $VE_{j} \in \left[0,1\right]$: Vaccine efficacy for agent $j$ (0 if unvaccinated, see \nameref{vaccination}).
\end{itemize}

Conceptually, Equation \ref{eq:transition} calculates the probability that agent $k$ is not infected by during a given time step, and is subtracted from 1 to yield the infection probability.
An agent’s actual transition is determined by drawing a random number $x \sim U(0,1)$; if $x < P_{k}$ then agent $k$ transitions to Exposed.

\textbf{Infection Attribution}
While EpiCast 2.0 does not deterministically identify the source of an infection (i.e. an individual source agent), it does track which venue infections are acquired within. Venue attribution is performed by tracking the total exposure an agent accumulates within each venue and then selecting a source venue by inverse transform sampling after an agent transitions to Exposed. Future versions of Epicast may be able to track the individual source agent.

\subsection{Policies} \label{policies}

Epicast 2.0 includes a flexible policy framework that enables users to simulate the impact of non-pharmaceutical interventions (NPIs), such as masking, distancing, work-from-home policies, and travel restrictions. These policies allow users to institute changes on specific days to the degree of participation in a venue (e.g. work or school) and the degree of social interactions within certain venues. Policies can be applied at the national, state, or county level. Policies can affect one of four domains: Work, School, Social interactions, and Travel.

Each policy is defined by:
\begin{enumerate}
    \item A time of enactment (as a relative day (e.g., day 40) or absolute date (e.g., 2020-04-05)),
    \item A domain of applicability (work, school, social, or travel),
    \item One or more selection specifiers (e.g., FIPS or NAICS codes),
    \item Domain-specific parameters (e.g., attendance rates or intensity scale factor).
\end{enumerate}

\subsubsection{Work policies}

Work policies control attendance and interaction intensity in workplace venues. The core parameters include:

\begin{itemize}
    \item \texttt{open}: Proportion of \textit{work-groups} that remain active (not individual workers).  For example, a work open proportion of 0.3 would mean that 70\% of work-groups would be \emph{closed} under that policy.
    \item \texttt{schedule}: Optional shift schedule (e.g., alternating weekdays). For example, a two-shift schedule could be constructed such that workers attend work on every other weekday (see \nameref{shift-schedules} for details).
    \item \texttt{scale}: A scalar in $[0, 1]$ modifying interaction intensity (IISF) to reflect interventions like distancing or masking. For example, a work IISF of 0.9 indicates a 10\% reduction in interaction intensity.
\end{itemize}

Work policies can target:

\begin{itemize}
    \item Specific locations, using county or state FIPS codes
    \item Specific industries, using 2- or 3-digit NAICS codes
\end{itemize}

If a location (industry) specifier is omitted from a policy, that policy will apply to all locations (industries). This can allow, for example, certain industries to be designated as \emph{essential} by specifying one policy that closes all industries and then another policy that opens only a few specific industries that the user might wish to designate as essential (policy precedence is given by user order). Furthermore, by using location selection specifiers, different states or counties can have different essential industries. This framework allows for a very high degree of policy specificity if desired. Note that users may define multiple overlapping policies. Later entries take precedence, allowing layered control (e.g., close all industries, then reopen only health care).

Listing \ref{lst:work_policy} shows an example of two work policies specified using TOML syntax, the first starting on simulation day 0 and simply opening all industries in all locations. The second, starting on April 13, 2020, is a compound policy that demonstrates how a user might specify location-specific essential industries. Note the use of two and three digit NAICS codes for controlling the degree of specificity in the industry selection specifiers. FIPS codes at the state and county level can be used in an analogous manner for location selection specifiers.

\begin{lstlisting}[language=TOML, caption=Example work policies, label={lst:work_policy}]
[0] # day 0, begin with all work-groups fully open
work = [ { open = 1.0 } ]

["2020-04-13"] # specify absolute date for policy to start in ISO 8601
work = [
    # all industries in all locations are closed
    { open = 0.0 },

    # in Arizona, all health care related industries (62X) are 100% open but with 
    # 50% reduction in interactions
    { open = 1.0, scale = 0.5, naics = [62], fips = [4] },

    # in New Mexico, hospitals (622) and live-in care facilities (623) are 100% 
    # open, all other health care industries (62X) are 50% open, interactions
    # for all health care industries are reduced by 40%
    { open = 0.5, scale = 0.6, naics = [62], fips = [35] },
    { open = 1.0, scale = 0.6, naics = [622, 623], fips = [35] }
]
\end{lstlisting}

\subsubsection{Shift schedules} \label{shift-schedules}

Shift schedules are 14 character long strings of the form: \lstinline[language=TOML]|"-AAAAA--AAAAA-"| where each character represents one day of a two week cycle, and alpha-numeric characters represent a \emph{shift} that is active on that day and non-alpha-numeric characters represent the abscense of an active shift. Day order is Sunday to Saturday, so the first character represents the first Sunday, and the last character represents the second Saturday, of each cycle. So \lstinline[language=TOML]|"-AAAAA--AAAAA-"| represents a typical five-days-per-week schedule with all agents attending as normal. Alternating daily shifts can be specified as: \lstinline[language=TOML]|"-ABABA--BABAB-"|, while alternating weekly shifts would be: \lstinline[language=TOML]|"-AAAAA--BBBBB-"|, where each day sees only 50\% of normal attendance. The main limitation of this framework is that only one shift can be active on each day, otherwise schedules can be quite complex.

\subsubsection{School policies}

School policies in Epicast 2.0 control in-person attendance behavior at different educational levels. Each policy defines two core parameters:

\begin{itemize}
    \item \texttt{attendance}: the proportion of students attending school in person, rather than the proportion of schools open
    \item \texttt{schedule}: an optional attendance rotation pattern (e.g., alternating days)
\end{itemize}

A value of $\texttt{attendance} = 0.8$ indicates $80\%$ of students attend physically, while the remaining 20\% participate remotely (i.e. stay home). Note that teachers are assumed to attend in person unless the school is fully closed (i.e., $\texttt{attendance} = 0$).

Like work policies, school attendance can also be structured using shifts (e.g., A/B group rotations), allowing alternating or hybrid schedules across days. These work analogously to work attendance schedules (see \nameref{shift-schedules} for details).

School policies may be applied selectively by: school-type and location. A school-type specifier is simply a list of the school types to which a school policy applies. As listed in \nameref{school-groups}, Epicast 2.0 school types include pre-kindergarten (pre-k), kindergarten, elementary, middle, and high schools. Location specifiers work exactly as they do for work policies, and can include state and county FIPS codes (or apply to all locations if omitted).

This setup allows users to simulate realistic, geographically targeted school closures or attendance modifications, for example, partial reopening of elementary schools in one state while high schools remain closed in another. Listing \ref{lst:school_policy} shows an example of two different ways to achieve approximately 50\% student attendance.  One option is by simply using a school attendance proportion of 0.5, and another is by using a rotating A/B schedule with $\texttt{attendance} = 1.0$, where all students attend on alternating days.

\begin{lstlisting}[language=TOML, caption=Example school policy, label=lst:school_policy]
[10] # policy starts on simulation day 10
school = [
    # state-wide, pre-k schools are closed
    { attendance = 0.0, school = ["pre-k"], fips = [35] },

    # state-wide, other schools are 50% in-person (so 50% of student always stay
    # home)
    { attendance = 0.5, school = ["kindergarten","elementary","middle","high"], fips = [35] },

    # except for Bernalillo County, which is using an every-other-day schedule
    # so also approximatly 50% overall attendance but all students attend half the time
    { attendance = 1.0, schedule = "-ABABA--BABAB-", school = ["middle","high"], fips = [35001] }
]
\end{lstlisting}

\subsubsection{Social and travel policies}

Social and travel policies in Epicast 2.0 allow users to modify interaction intensity in public settings and restrict long-distance movement. These policies reflect the types of broad public health guidance often issued at the state or county level.

The \emph{social domain} refers to all public, multi-household interaction venues (community, neighborhood, household-cluster, school) where county or state level public health recommendations around physical distancing and masking are most relevant. Note that the work domain has a separate IISF, as companies might enact -- for example -- stricter or looser policies than the county or state. Currently, Epicast 2.0 does not have a separate IISF for schools, instead the social domain IISF applies to schools.

Travel policies control the probability of irregular (long-distance) travel originating from specific geographic areas. The key parameter is a trip likelihood scale factor in $[0, 1]$, which scales the base probability that agents in a given state or county will initiate an irregular trip. A value of 0.0 halts all irregular travel from the specified area.

Both social and travel policies use location-based specifiers (i.e., lists of FIPS codes) and are structured identically to work and school policies. Examples of social and travel policies are given in Listing~\ref{lst:soc_trav_policy}.

\begin{lstlisting}[language=TOML, caption=Example social and travel policies, label={lst:soc_trav_policy} ]
["2020-04-13"]
social = [
    # no state-wide distancing or masking in New Mexico
    { scale = 1.0, fips = [35] },

    # stricter policies in effect in Bernalillo and Santa Fe counties
    { scale = 0.5, fips = [35001, 35049] },

    # 20% state-wide reduction in interaction intensity in Arizona
    { scale = 0.8, fips = [4] }
]

travel = [ { scale = 0.0 }] # all irregular / long distance travel is stopped
\end{lstlisting}

\subsection{Other initial condition data}

\subsubsection{Index cases and initial immunity} \label{index-cases}

The number of index cases (agents that start in the Exposed compartment) and the size of the initially immune population (agents that start in the Recovered compartment) per county can be specified in one of two ways. The first is as a simple, three column table of the form show in Table \ref{tab:index_cases}.

\begin{table}[h]
    \begin{center}
        \begin{tabular}{|c|c|c|}
            \hline
            \thead{County FIPS} & \thead{\# index cases} & \thead{\# immune} \\
    	    \hline
            \ldots & \ldots & \ldots \\
            \hline
    	\end{tabular}
        \caption{Format for manually specifying the number of index cases and initially immune agents per-county.}
        \label{tab:index_cases}
    \end{center}
\end{table}

If a user supplied table is not given, the number of index cases and immune population for each county within the simulation is calculated from a database of daily new and cumulative cases of SARS-CoV-2. The database includes case counts for each county in the US spanning January 21, 2020 to March 23, 2023, and was generated from data aggregated by The New York Times based on reports from state and local health agencies \cite{NytimesCovid19data2023}. Users can also specify scale factors for the number of index cases and the size of the initially immune population that can, for example, approximate under-reporting in case count data.

Initial immunity is assigned before index case assignment takes place and is performed such that every agent within the same county has an equal probability of starting out immune (i.e. assignment is random). Index cases are assigned in a similar manner, except that only susceptible agents are considered, so all susceptible agents within the same county have an equal probability of being moved to the Exposed compartment at the time of assignment. This is important, as in Epicast 2.0 index case seeding is performed on each of the first $N$ days of the simulation, where $N$ is a user specified index case multiplier. Often $N$ is set to \alttilde 5. Therefore, the number of susceptible agents can change significantly over the $N$ days of index case seeding and so must be taken into account when making index case assignments. 

\subsubsection{Prophylactic and therapeutic interventions} \label{vaccination}

Within Epicast 2.0, vaccination is the only supported pharmaceutical intervention. Agents can be in one of three vaccination states: unvaccinated, partially vaccinated, and fully vaccinated. Vaccines reduce: the degree of shedding of infectious agents, the probability that Susceptible agents will transition to Exposed, the probability that Pre-symptomatic agents will transition to Symptomatic, and the duration of the Symptomatic and Asymptomatic states by one day. Table \ref{tab:vaccine_scalars} lists the default scale factors by which each level of vaccination modifies probabilities / likelihoods within Epicast 2.0.

\begin{table}[h]
    \begin{center}
        \begin{tabular}{@{}lcccc@{}}
            \toprule
            \thead{} & \thead{Degree of shedding} & \thead{$P(S \rightarrow E)$} & \thead{$P(I_{P} \rightarrow I_{S})$} \\
    	    \midrule
            Partially vaccinated & 0.8 & 0.7 &  1.0 \\
            Fully vaccinated & 0.8 & 0.7 & 0.6 \\
    	\end{tabular}
        \caption{Effect of partial and full vaccination on agent shedding (column one) and agent transitions (columns two and three). $P(S \rightarrow E)$: probability of transitioning from Susceptible to Exposed; $P(I_{P} \rightarrow I_{S})$: probability of transitioning from Pre-symptomatic to Symptomatic. }
        \label{tab:vaccine_scalars}
    \end{center}
\end{table}

\subsubsection{Work-group size} \label{work-group-data}

\begin{figure}[h]
    \centering
    \includegraphics[width=16.5cm]{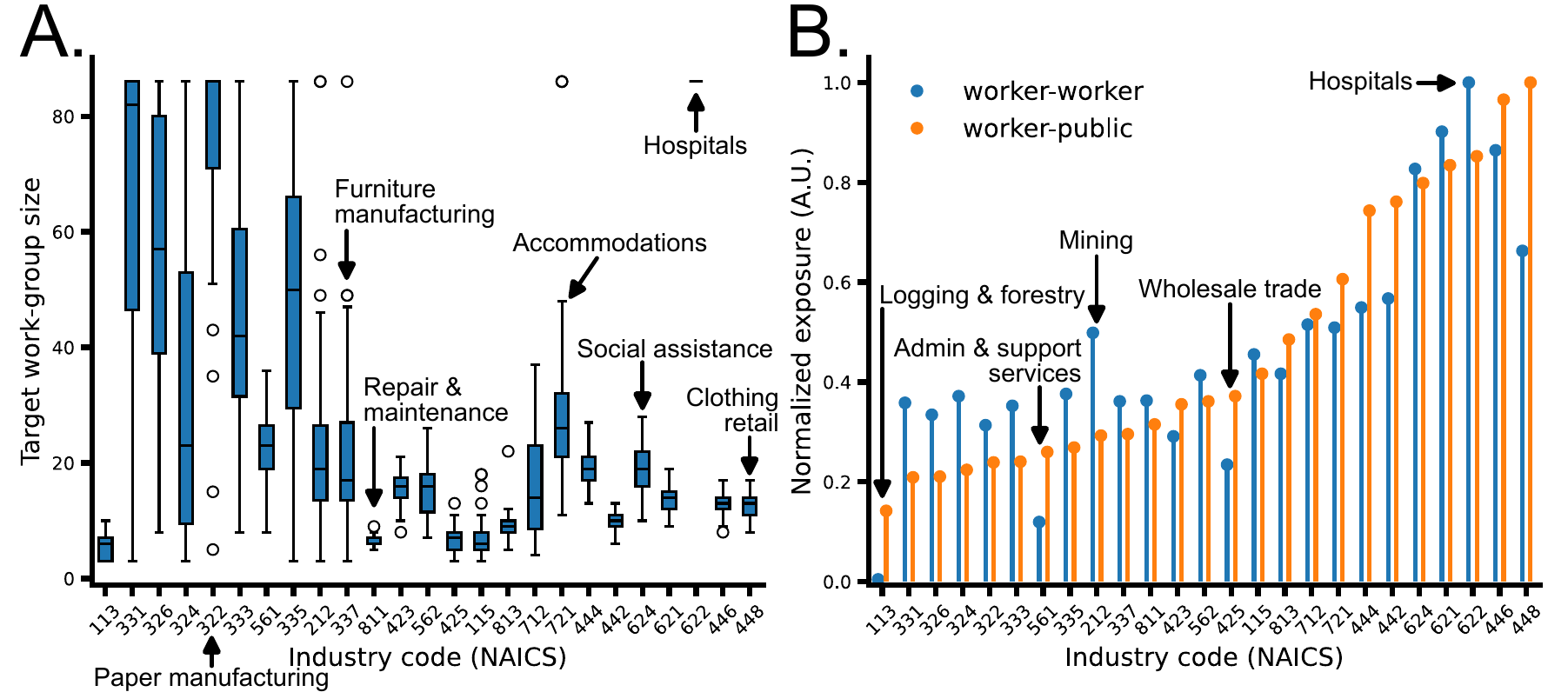}
    \caption{Summary of work-group size and worker contact intensity data. \textbf{A}) Work-group size data for 25 selected industries indicated by their NAICS code. Boxplots show the variation across states within industry. Work-groups are limited to 86 agents at the largest. This limit effects some manufacturing industries (NAICS codes in the 200-400 range), but most prominently hospitals (NAICS 622) where the largest work-groups are seen. \textbf{B}) Worker-worker IISF (blue) and worker-public IISF (orange) for the same select industries as shown in \textbf{B}. For visual clarity, both types of IISF are scaled to $\left[0,1\right]$. See \nameref{sec:worker_contact_intensity} for details.}
    \label{fig:work_data}
\end{figure}

Industry and US state specific work-group sizes were derived from the 2019 County Business Patterns survey from the US Census Bureau (CPB) \cite{CountyBusinessPatterns2019}. We limited the maximum work-group size to 86 based on studies of workplace contact patterns \cite{lowery-northMeasuringSocialContacts2013}. The CBP data provide a \emph{target work-group size} for each industry (3-digit NAICS code) in each US state that the work-group assignment algorithm uses in determining the number of work-groups that each industry in each community requires. The distributions of target work-group sizes for 25 example industries are shown in Figure \ref{fig:work_data}A. Given that the assignment of workers to work-groups is random within each industry, the average work-groups size for each industry will approximately match the CBP data, but with some variation (see \nameref{work-group-assign}).

\subsubsection{Worker contact intensity} \label{sec:worker_contact_intensity}

Given that workers in different industries are likely to experience varying levels of interactions (in both number and intensity), Epicast 2.0 includes the option for the user to specify industry specific IISF for both worker-worker and worker-public interactions. By default, these IISF are uniform across industries; however, Epicast 2.0's auxiliary data includes an example data set of work related IISF derived from data from the Occupational Information Network (O*NET) of the US Department of Labor \cite{ONETOnline2024} and analyses thereof by Avdiu and Nayyar \cite{avdiuWhenFacetofaceInteractions2020}.

In the worker contact data, each industry (3-digit NAICS code) $i$ that occurs within the UrbanPop population has two IIS, one for worker-worker interactions, $\beta_{i}$, and one for worker-public interactions, $\alpha_{i}$. $\beta_{i}$ is the O*NET score resulting from the question: ``To what extent does this job require the worker to perform job tasks in close physical proximity to other people?'' \cite{ONETOnline2024} scaled to $\left[0.33,1.0\right]$ across industries. $\alpha_{i}$ is just $\beta_{i} \times c_{i}$ scaled to $\left[0.0033,0.01\right]$ across industries, where $c_{i}$ is the Face-to-Face public interaction score for industry $i$ calculated by Avdiu and Nayyar \cite{avdiuWhenFacetofaceInteractions2020}. The maximum worker-public IISF of 0.01 accounts for the large number of public-worker interactions (up to \alttilde 2000 per worker) that occur within Epicast 2.0's all-to-all interaction model (as every agent in a community is a member of the public in this context). Figure \ref{fig:work_data}B shows the worker-worker and worker-public IISF for select industries. Note that the IISF are shown scaled to $\left[0,1\right]$ for visual clarity, actual IISF are scaled as described above.

In terms of Equation \ref{eq:exposure}, $\beta_{i}$ simply scales the $C_{v}$ contact matrix for the work-group venue. $\alpha_{i}$ can be thought of as effectively introducing a new venue in which workers interact with the public and each element of $C_{v}$ for that venue is set to $\alpha_{i}$.

\subsubsection{School-group size} \label{school-group-data}

Target school-group sizes for each school type and each county were derived from student-teacher ratio data from the National Center for Education Statistics Elementary/Secondary School Information System \cite{ElementarySecondaryInformation2019}. School-groups were limited to 50 students as a best-guess maximum (state average class sizes rarely exceed 30 students \cite{NationalTeacherPrincipal2020}). The school-group size data play an analogous role to work-group size data in providing a target group size for the assignment algorithm, but with higher spatial resolution (county vs. state).

\subsection{Implementation notes}

Epicast 2.0 is implemented in the C programming language and depends only on the C standard library and an implementation of the Message Passing Interface (MPI) version 3.0 or greater \cite{MPIMessagePassingInterface2015}. All user supplied parameters can be specified on the command line or via a configuration file using the TOML syntax.

\section{Demonstration scenarios}

We demonstrate the functionality described above in the following scenarios that are meant to highlight a few of the core features of Epicast 2.0. First, we show examples of relatively long simulations (1.5 years) of the contiguous US (lower 48 states and Washington DC), initialized with observed case data from the SARS-CoV-2 pandemic, that highlight the effect of varying the main transmissibility parameter $P_\text{trans}$ on the resulting outbreak dynamics. Next, we highlight the spatial aspects of pathogen spread by seeding index cases within a single county of the northeastern US and show how the pathogen spreads though the rest of the region over time. Finally, we demonstrate the use and effects of polices on outbreak dynamics using New Mexico as a case study.

\subsection{Varying transmissibility}

\begin{figure}[h]
    \centering
    \includegraphics[width=16.5cm]{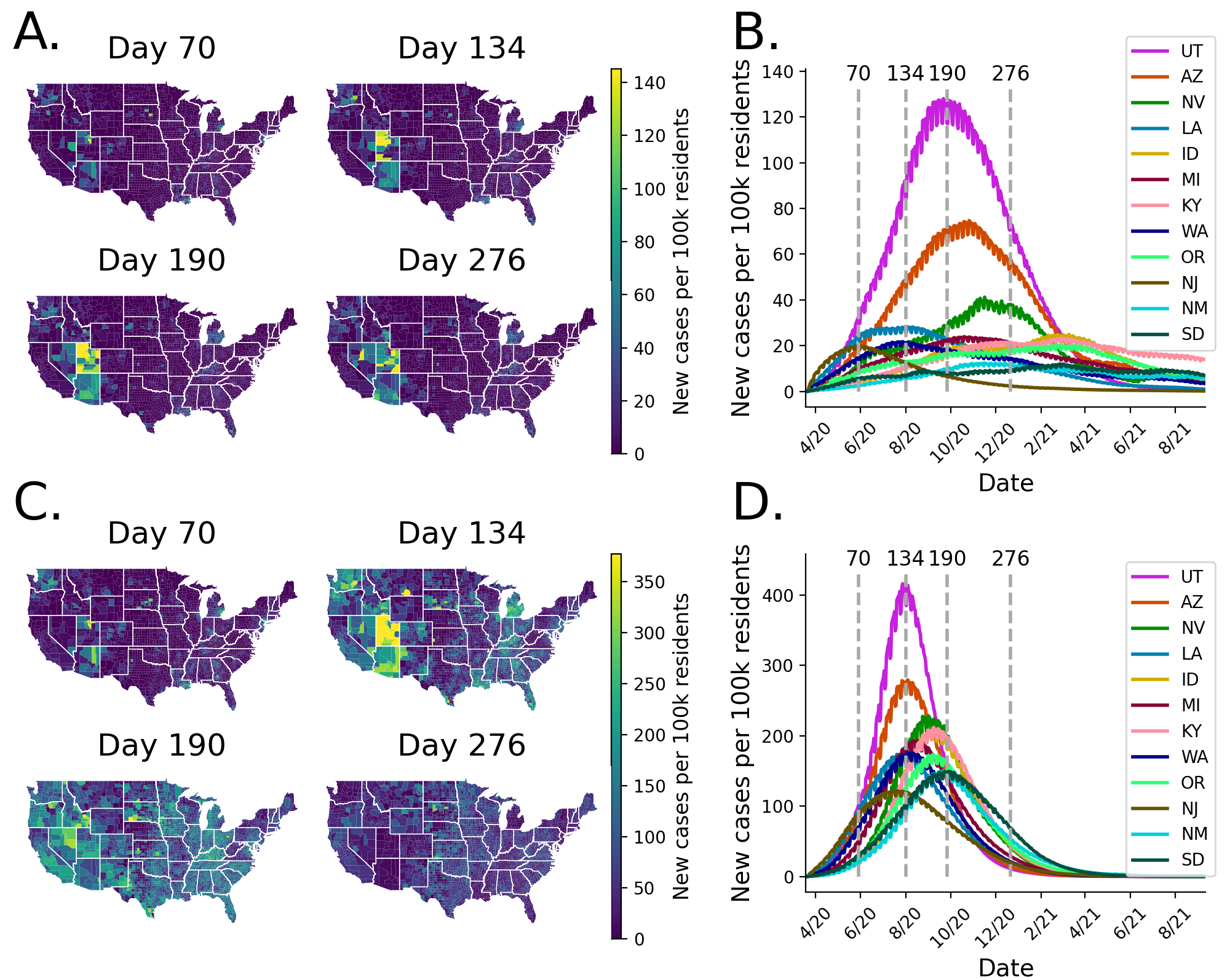}
    \caption{Example simulations of the lower-48 US states and Washington DC starting on March 20, 2020 and ending on September 11, 2021 (540 days). \textbf{A} and \textbf{B} (\textbf{C} and \textbf{D}) illustrate a simulation that used a $P_\text{trans}$ of 0.06 (0.08). Index cases were seeded on the first five days of each simulation, with the number and spatial distribution given by the observed SARS-CoV-2 case data from March 25, 2020. \textbf{A} and \textbf{C} show the spatial distribution of new cases (per 100,000 residents) in each county on four days during the outbreak. \textbf{B} and \textbf{D} show the time course of new cases for the 12 states with the highest peak in each simulation. Vertical, dashed grey lines indicate the time points shown in \textbf{A} and \textbf{C}.}
    \label{fig:full_us_ptrans}
\end{figure}

Figure \ref{fig:full_us_ptrans} shows how varying the main transmissibility parameter, $P_\text{trans}$, affects the outbreak dynamics. Panels A and B show the results of a simulation with $P_\text{trans}$ set to 0.06 and panels C and D show the same outbreak measures but for a $P_\text{trans}$ of 0.08. Both simulations began on March 20, 2020 and each county was seeded with its observed SARS-CoV-2 case count from March 25, 2020 (index cases were seeded on each of the first five days). The number of initially immune agents in each county was set to four times the observed cumulative case load through March 25 (approximating a 4x under-reporting of cases in the period prior to March 25). All parameters, except $P_\text{trans}$, were identical between the two simulations and no mitigation strategies were used (i.e. agents participated in work, school, and irregular travel as usual throughout the simulation).

As expected, a higher $P_\text{trans}$ resulted in higher case counts, with 4.5\% of agents infected when $P_\text{trans}$ was set to 0.6 (Figure \ref{fig:full_us_ptrans}A and B) and 24.8\% of agents infected when $P_\text{trans}$ was set to 0.8 (Figure \ref{fig:full_us_ptrans}C and D). Similarly, the spatial distribution of cases is much broader with a higher $P_\text{trans}$, especially during the peak of the outbreak (compare Figure \ref{fig:full_us_ptrans}A and Figure \ref{fig:full_us_ptrans}C for days 134 and 190). The change in temporal dynamics with increasing $P_\text{trans}$ was more pronounced. At the higher transmissibility, state-level outbreaks are approximately Gaussian shaped and vary mostly in peak amplitude (Figure \ref{fig:full_us_ptrans}D), while at the lower transmissibility state-level outbreaks are non-Gaussian and often asymmetric (e.g. Nevada, green line in Figure \ref{fig:full_us_ptrans}B), and the outbreaks in some states haven't fully subsided within the 540-day simulation window (most notably Kentucky, light-pink line in Figure \ref{fig:full_us_ptrans}B).

\subsection{Seeding index cases in a single county}

\begin{figure}[h]
    \centering
    \includegraphics[width=16.5cm]{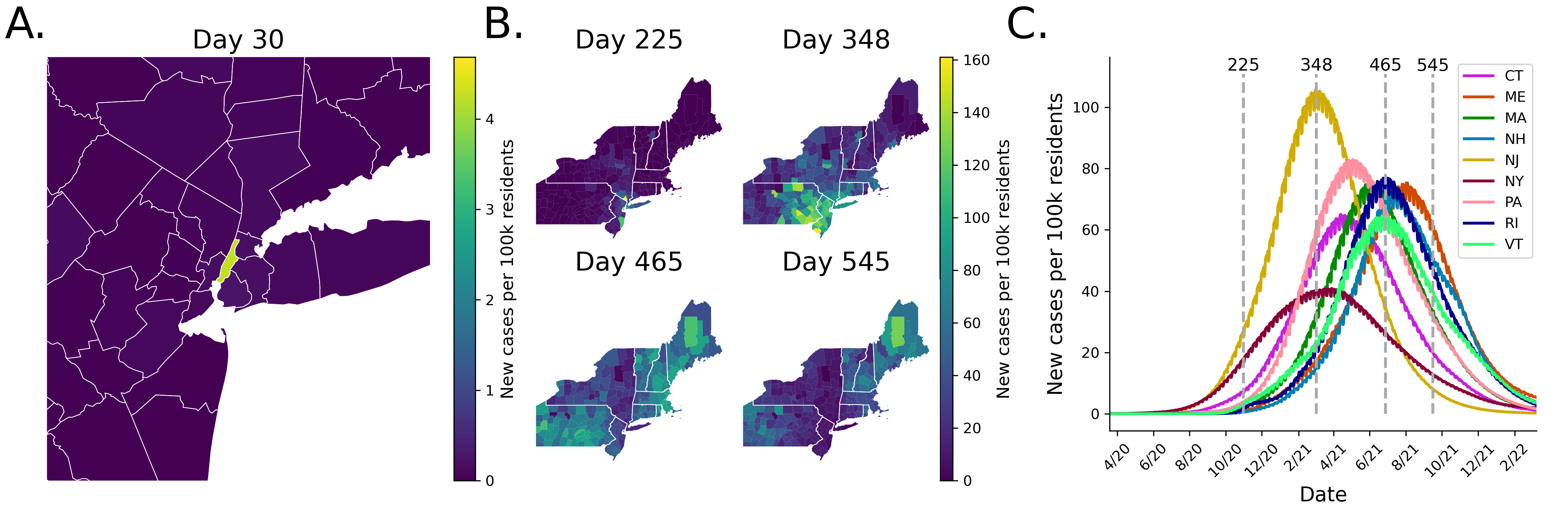}
    \caption{Example 720-day simulation of the northeast US where 1000 index cases were seeded in New York County (200 cases on each of the first five days). \textbf{A}) Spatial distributions of cases on day 30, highlighting the location of New York County where all 1000 index cases were seeded. \textbf{B}) Spatial distribution of new cases (per 100,000 residents) for all counties within each state on four example days of the outbreak. \textbf{C}) Time course of new cases for the 9 states included in the simulation. Dashed grey lines indicate the timing of the days highlighted in \textbf{B}.}
    \label{fig:ne_1county_seed}
\end{figure}

Figure \ref{fig:ne_1county_seed} shows the results of a 720-day simulation of the northeastern US where 1000 index cases were seeded within New York County. All other counties started with zero index cases. Through the first 30 days, the outbreak remains almost entirely confined within New York County as shown in Panel A. Panel B shows how the spatial distribution of new cases evolves over longer time scales. By day 225 of the simulation (Figure \ref{fig:ne_1county_seed}B, upper left) the outbreak is still largely confined to southern New York state and northern New Jersey, with smaller outbreaks beginning to emerge in central New York state. By day 348 (Figure \ref{fig:ne_1county_seed}B, upper right) the outbreak has spread considerably, covering most of New Jersey and the eastern half of Pennsylvania. Panel C shows that case counts in both New York (dark-red line in Figure \ref{fig:ne_1county_seed}C) and New Jersey (yellow) are peaking at this time, while all other states are experiencing a rapid rise in cases, most notably Pennsylvania (pink), Connecticut (magenta) and Massachusetts (green). By day 465 (Figure \ref{fig:ne_1county_seed}B, lower left) the outbreak is waning rapidly in New York and New Jersey, while case counts are peaking in Rhode Island (Figure \ref{fig:ne_1county_seed}C dark-blue) and Vermont (light-green). By day 545 the outbreak is subsiding in all states, with New Hampshire (blue) and Maine (orange) still somewhat elevated having been the last states to peak.

Note that this simulation used the same value of $P_\text{trans}$ (0.08) as the simulation illustrated in Figure \ref{fig:full_us_ptrans}C and D. Visual comparison of the state-level time courses (Figure \ref{fig:full_us_ptrans}D versus \ref{fig:ne_1county_seed}C) highlights the extent to which, under default settings, a very local (i.e. single county) outbreak takes a long time (\alttilde 200 days) to gain momentum within the model, even when the population outside of the seed county is completely susceptible.

\subsection{Effect of policies on outbreak dynamics}

\begin{figure}[h]
    \centering
    \includegraphics[width=16.5cm]{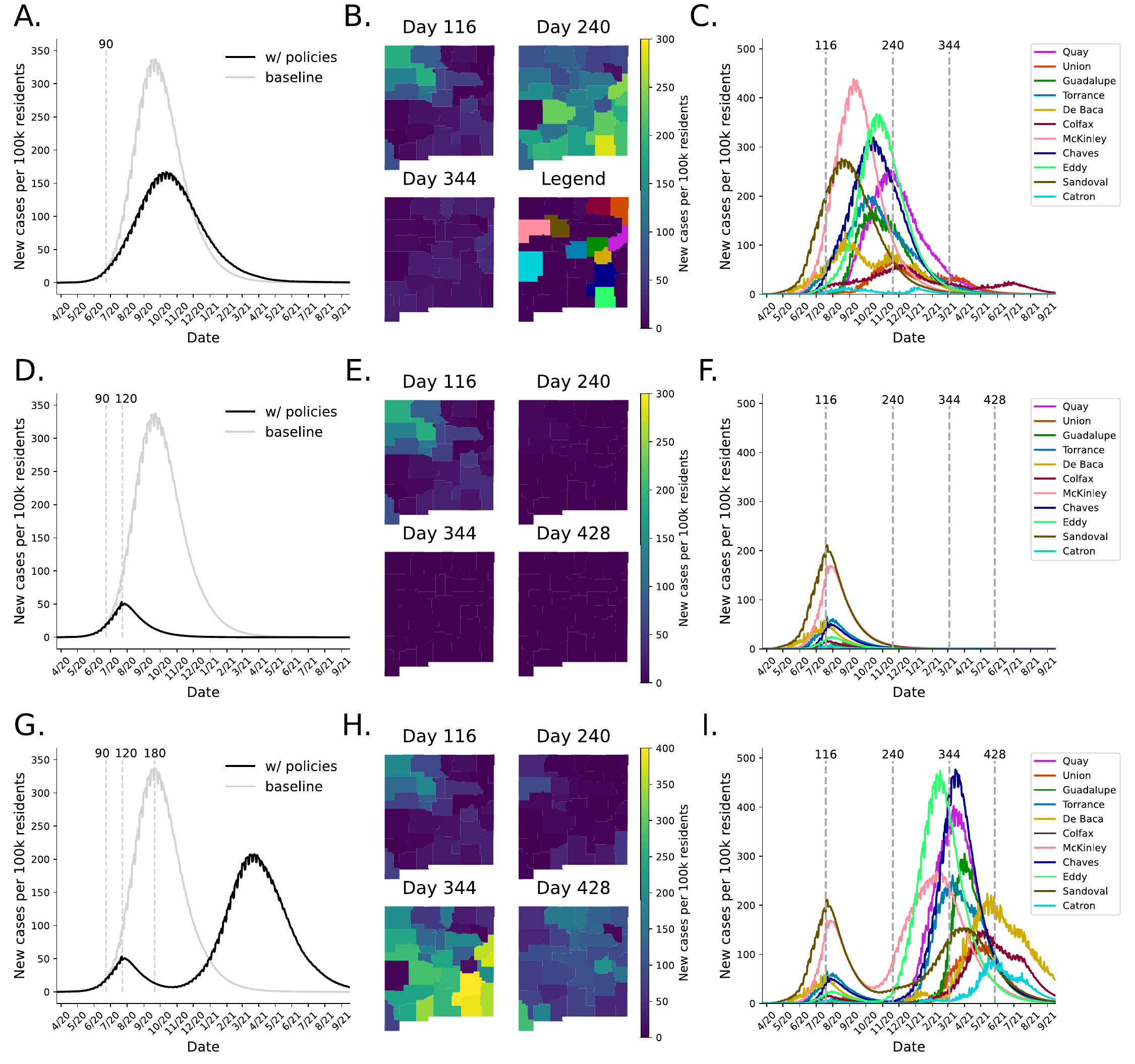}
    \caption{Demonstration of the effect of policies on an example, 540-day outbreak within New Mexico. Rows represent three different sets of policies applied to the same outbreak that was seeded with observed SARS-CoV-2 case counts from March 25, 2020. \textbf{A-C}) Spatiotemporal pattern of cases resulting from the application of a single, moderately restrictive policy enacted on day 90. \textbf{D-F}) Same as \textbf{A-C} but with an additional, more restrictive policy enacted on day 120. \textbf{G-I}) Same as \textbf{D-F} but with a loosening of restrictions on day 180. Solid black lines in \textbf{A,D,G} show the state-level case timeseries with the corresponding policies (dashed vertical lines) in place. Solid grey lines show what the state-level case timeseries would have been in the absence of any policies (baseline). \textbf{B,E,H} show the spatial distribution of cases on select days of the outbreak. The lower right sub-panel in \textbf{B} (labeled \protect\say{Legend}) shows the locations (via corresponding colors) of the 11 counties whose timeseries are shown in \textbf{C,F,I}.}
    \label{fig:nm_policy}
\end{figure}

Figure \ref{fig:nm_policy} shows the results of three simulations in which a series of polices are successively applied, highlighting the effect of each policy. All (non-policy) simulation parameters and random number generator seeds were identical, ensuring that any difference between simulations is exclusively the result of differences in the policies that were applied (e.g. the outbreaks illustrated in Figure \ref{fig:nm_policy} A-C and D-F are identical up to day 120). Column one (Figure \ref{fig:nm_policy}A,D,G) shows the state-level case timeseries (solid black lines) that are produced when the indicated policies (vertical dashed lines) are applied as well as the case timeseries that would have been produced had no policies been enacted (i.e. baseline condition, solid grey lines). Column two shows the spatial distribution of new cases on four days during each outbreak. Note that the lower-right sub-panel in Figure \ref{fig:nm_policy}B is a legend, indicating the location of each county with colors that correspond to the county-level case timeseries shown in column three.

These scenarios are meant to illustrate an outbreak response in which a moderately restrictive policy is implemented just as the outbreak is starting to take off (day 90) followed by a more restrictive policy 30 days later (day 120) as daily case counts continue to grow despite the restrictions. As daily new cases fall to very low levels (\alttilde 15 new cases per 100,000 residents by day 180), all restrictions are then lifted. The moderate policy is implemented as a 25\% reduction in interaction intensity within work, school, neighborhood, and community venues approximating the effect of occasional masking and physical distancing without any change in school or work attendance. The more restrictive policy closes all schools, closes 50\% of businesses, and implements a 50\% reduction in interaction intensity within the work, neighborhood, and community venues. The exact specification of the three policies are shown in Listing \ref{lst:nm_policy}.

\begin{lstlisting}[language=TOML, caption=Policies used for simulations shown in Figure \ref{fig:nm_policy}, label={lst:nm_policy} ]
[90]
work = [ {open = 1.0, scale = 0.75} ]
social = [ {scale = 0.75} ]
travel = [ {scale = 0.75} ]

[120]
work = [ {open = 0.5, scale = 0.5} ]
school = [ {open = 0.0} ]
social = [ {scale = 0.5} ]
travel = [ {scale = 0.5} ]

[180]
work = [ {open = 1.0, scale = 1.0} ]
school = [ {open = 1.0} ]
social = [ {scale = 1.0} ]
travel = [ {scale = 1.0} ]
\end{lstlisting}

Figure \ref{fig:nm_policy}A-C show how the outbreak would unfold if only the first, moderately restrictive, policy was enacted. Clearly, far fewer cases occur than in the baseline condition (Figure \ref{fig:nm_policy}A), but daily case counts still exceed 200 per 100,000 residents in several counties and reach \alttilde 150 state wide (Figure \ref{fig:nm_policy}A \& C). For reference, during the winter of 2020-21 SARS-CoV-2 cases in New Mexico peaked at \alttilde 120 per 100,000 residents. Only during the winter peak of 2021-22 did cases exceeded 150 per 100,000 residents state wide (reaching \alttilde 330 at the end of January 2022, similar to the baseline condition simulated here) \cite{NytimesCovid19data2023}. Figure \ref{fig:nm_policy}D-F show how the outbreak would unfold if a second, more restrictive, policy was put into place on day 120. In this case, state-wide daily cases barely surpass \alttilde 50 per 100,000 (Figure \ref{fig:nm_policy}D), and only two counties see peak case counts over \alttilde 100 (Figure \ref{fig:nm_policy}F). Finally, Figure \ref{fig:nm_policy}G-I show the full scenario where all restrictions are lifted on day 180 when daily, state-level new cases fall below \alttilde 15 per 100,000 residents. In this case, even though state-level cases (as well as cases in all but two counties) fall as low as \alttilde 7 per 100,000 (around day 200), a major second wave occurs around 6 months after all restrictions are lifted. While this second wave peaks at a substantially lower daily case load than the peak of the baseline condition (\alttilde 330 vs. \alttilde 200, Figure \ref{fig:nm_policy}G), the overall proportion of infected agents is only modestly lower (41.4\% baseline vs. 36.1\% full scenario). For comparison, the overall proportion of infected agents is 29.9\% and 7.2\% for the one policy (Figure \ref{fig:nm_policy}A-C) and two policy (Figure \ref{fig:nm_policy}D-F) scenarios, respectively.

\section{Discussion}

Epicast 2.0 introduces a number of significant updates to better reflect the structural, behavioral, and demographic heterogeneity of the United States. By incorporating a detailed synthetic population from UrbanPop \cite{tuccilloUrbanPopSpatialMicrosimulation2023}, Epicast 2.0 enables the construction of contact networks that are more representative of real-world patterns—particularly in schools, workplaces, and local communities.

The updated network and transmission models draw on a diverse set of data sources, including industry-specific contact rates and empirical group size distributions. These inputs allow for finer-grained simulations of transmission risk across occupations and educational settings. Epicast 2.0 also includes a highly flexible policy framework that supports time-varying interventions targeted at specific school types, industries, and geographic regions.

Taken together, these improvements support realistic scenario modeling and counterfactual exploration -- key tools for informing future public health policy. Importantly, the model can simulate how different communities might respond to the same policy differently, due to underlying differences in population structure, mobility, and contact patterns.

\subsection{Principal application domain}

The primary use case for Epicast 2.0 is to explore counterfactual scenarios in a high-fidelity, data-driven simulation environment. While empirical data can inform the evaluation of past policies, they do not reveal what might have happened under alternative decisions. Simulation is therefore essential for understanding the comparative effectiveness of unimplemented or hypothetical interventions.

Crucially, model fidelity at national scale requires not only population heterogeneity, but also realistic correlations between attributes -- such as household size, occupation, and geography -- that influence exposure and vulnerability. The UrbanPop-generated synthetic population captures these correlations, supporting simulations that better reflect the uneven burden of disease observed in real epidemics, including SARS-CoV-2 and influenza \cite{quinnRacialDisparitiesExposure2011, bassettVariationRacialEthnic2020, fergusonGeographicTemporalVariation2022, irizarEthnicInequalitiesCOVID192023}. The combination of the UrbanPop synthetic population and the improved network construction system that Epicast~2.0 employs produces transmission dynamics that can recapitulate the observed heterogeneity in the distribution of SARS-CoV-2 cases.

The combination of demographic realism, venue-specific transmission dynamics, and flexible policy implementation makes Epicast 2.0 a powerful platform for simulating diverse “what-if” scenarios. For example, simulations could be calibrated to observed case data for a specific region (e.g., New Mexico) under historical policies, then rerun with alternative interventions to estimate counterfactual outcomes.

\subsection{Limitations and future work}

As discussed in Section~\ref{sec:epicast_framework}, the primary limitations of the Epicast framework are: 
(1) agent interactions occur only within communities during each time step, and 
(2) interactions within a venue are all-to-all. These approximations significantly improve computational performance and make national-scale simulations tractable, but they introduce trade-offs in spatial and network realism.

\vspace{1em}
\noindent\textbf{Community isolation}  
By design, Epicast restricts interactions to within-community mixing for each time step. This assumption limits the model's ability to represent high-traffic venues or gathering places (e.g., malls, large workplaces) that draw individuals from multiple communities. While the resulting over-compartmentalization is likely to average out at large spatial scales (e.g., state-level), it may reduce fidelity at the census tract or county level. Future work will incorporate mechanisms to allow more frequent travel to adjacent communities during designated ``community time'' periods (e.g., weekends and holidays) to better capture these effects.

\vspace{1em}
\noindent\textbf{All-to-all mixing within venues}  
Epicast models each venue as a fully connected contact network. While efficient, this can overestimate exposure when many infectious agents are present and under-represent realistic transmission chains. The impact is greatest at small scales, where stochasticity in exposure is amplified. Nonetheless, the all-to-all model is conceptually equivalent to an "accumulated exposure" framework, where the order and identity of contacts are not tracked. Given Epicast’s focus on scenario and policy modeling, this approximation is typically acceptable and enables fast exploration of complex intervention spaces.

\vspace{1em}
\noindent\textbf{Regular travel model limitations}  
Epicast's commute model is based on unconditioned tract-to-tract worker flow data. This can lead to unrealistic assignments -- most notably, teachers may be over-allocated to workplace-dense tracts without schools. Future updates will incorporate industry-conditioned commute data to ensure that, for example, educators are more likely to be assigned to school-serving tracts. Additionally, students are currently restricted to attending school within their home community. Allowing cross-community student commutes will better represent realistic school catchment areas and enhance inter-community mixing in a computationally efficient manner.

Other areas of active or planned development include: enhancing the irregular travel model to incorporate seasonality, purpose of travel, and real-world travel data; supporting dynamic or reactive policies that depend on case counts or other model-derived metrics; incorporating bottom-up behavioral models alongside top-down policy levers; expanding capabilities to model multiple concurrent pathogens; introducing agent-specific variation in compliance, risk perception, and responsiveness.

These improvements will enhance Epicast’s fidelity at local scales, expand its relevance across pathogens and contexts, and increase its value as a decision-support tool for policy makers and public health agencies.

\subsection{Conclusions}

Epicast~2.0 is a national-scale, agent-based modeling platform for simulating the spread of respiratory pathogens within the United States. It incorporates a demographically detailed synthetic population of approximately 324 million agents, constructed using high-fidelity US Census data from 2019 \cite{tuccilloUrbanPopSpatialMicrosimulation2023}. These agents are organized into households, schools, workplaces, neighborhoods, and communities using empirical data on group sizes, mobility, and employment structure.

Epicast~2.0 supports a wide array of user-defined non-pharmaceutical interventions, including policies that alter school and work attendance, regulate physical distancing, and restrict long-distance travel. Policies can be enacted at the national, state, or county level, and can vary by industry, school type, or community attributes. The framework also includes realistic travel models, age-structured contact intensities, and pathogen-specific transmission dynamics. The model is optimized for high-performance computing environments, enabling national-scale simulations while retaining the ability to analyze outcomes at the individual agent level.

Although Epicast~2.0 was originally developed to simulate the spread of SARS-CoV-2, its pathogen-agnostic structure makes it well-suited for modeling a wide range of respiratory diseases. It can be used both retrospectively, to analyze past pandemic responses, and prospectively, to evaluate counterfactual or future mitigation strategies.

\vspace{1em}
\noindent\textbf{Policy Implications}
Epicast~2.0 provides an important tool for supporting public health decision-making. It enables:

\begin{itemize}
    \item \textit{Counterfactual policy evaluation}, by simulating interventions that were not implemented in reality;
    \item \textit{Scenario planning and stress testing}, under different transmissibility, compliance, or vaccination assumptions;
    \item \textit{Demographic heterogeneity}, through demographic detail that allows exploration of differences in disease burden across age, race, ethnicity, and socioeconomic status;
    \item \textit{Targeted intervention modeling}, including layered NPIs and phased reopening strategies at fine spatial scales.
\end{itemize}

By grounding policy evaluation in realistic agent behavior and venue-based interactions, Epicast~2.0 bridges the gap between high-level epidemic modeling and operational decision-making.

\section{Acknowledgments}

This work was supported by the US Department of Energy, Office of Science, Office of Advanced Scientific Computing Research and by cooperative agreement CDC-RFA-FT-23-0069 from the CDC’s Center for Forecasting and Outbreak Analytics. Its contents are solely the responsibility of the authors and do not necessarily represent the official views of the Centers for Disease Control and Prevention.

This work was performed at Los Alamos National Laboratory (LANL), an equal opportunity employer, which is operated by Triad National Security, LLC, for the National Nuclear Security Administration (NNSA) of the US Department of Energy (DOE) under contract \#19FED1916814CKC. The funders had no role in study design, data collection and analysis, decision to publish, or preparation of the manuscript. This work is approved for public distribution under LA-UR-25-23225. The findings and conclusions in this report are those of the authors and do not necessarily represent the official position of LANL.

This research used resources provided by the Darwin testbed at LANL which is funded by the Computational Systems and Software Environments subprogram of LANL's Advanced Simulation and Computing program (NNSA/DOE).

\bibliographystyle{vancouver}
\bibliography{references}

\begin{thebibliography}{10}

\bibitem{rayEnsembleForecastsCoronavirus2020}
Ray EL, Wattanachit N, Niemi J, Kanji AH, House K, Cramer EY, et~al.. Ensemble
  {{Forecasts}} of {{Coronavirus Disease}} 2019 ({{COVID-19}}) in the
  {{U}}.{{S}}.. medRxiv; 2020.
\newblock Available from:
  \url{https://www.medrxiv.org/content/10.1101/2020.08.19.20177493v1}.

\bibitem{walkerImpactCOVID19Strategies2020}
Walker PGT, Whittaker C, Watson OJ, Baguelin M, Winskill P, Hamlet A, et~al.
\newblock The Impact of {{COVID-19}} and Strategies for Mitigation and
  Suppression in Low- and Middle-Income Countries.
\newblock Science. 2020 Jul;369(6502):413-22.
\newblock Available from:
  \url{https://www.science.org/doi/10.1126/science.abc0035}.

\bibitem{cramerUnitedStatesCOVID192022}
Cramer EY, Huang Y, Wang Y, Ray EL, Cornell M, Bracher J, et~al.
\newblock The {{United States COVID-19 Forecast Hub}} Dataset.
\newblock Scientific Data. 2022 Aug;9(1):462.
\newblock Available from:
  \url{https://www.nature.com/articles/s41597-022-01517-w}.

\bibitem{howertonEvaluationUSCOVID192023}
Howerton E, Contamin L, Mullany LC, Qin M, Reich NG, Bents S, et~al.
\newblock Evaluation of the {{US COVID-19 Scenario Modeling Hub}} for Informing
  Pandemic Response under Uncertainty.
\newblock Nature Communications. 2023 Nov;14(1):7260.
\newblock Available from:
  \url{https://www.nature.com/articles/s41467-023-42680-x}.

\bibitem{tracyAgentBasedModelingPublic2018}
Tracy M, Cerd{\'a} M, Keyes KM.
\newblock Agent-{{Based Modeling}} in {{Public Health}}: {{Current
  Applications}} and {{Future Directions}}.
\newblock Annual Review of Public Health. 2018 Apr;39(Volume 39, 2018):77-94.
\newblock Available from:
  \url{https://www.annualreviews.org/content/journals/10.1146/annurev-publhealth-040617-014317}.

\bibitem{steinhofelAgentBasedModelsVirus2025}
Steinh{\"o}fel KK, Heslop D, MacIntyre CR.
\newblock Agent-{{Based Models}} of {{Virus Infection}}.
\newblock Current Clinical Microbiology Reports. 2025 Jan;12(1):2.
\newblock Available from: \url{https://doi.org/10.1007/s40588-024-00238-5}.

\bibitem{germannMitigationStrategiesPandemic2006}
Germann TC, Kadau K, Longini IM, Macken CA.
\newblock Mitigation Strategies for Pandemic Influenza in the {{United
  States}}.
\newblock Proceedings of the National Academy of Sciences. 2006
  Apr;103(15):5935-40.
\newblock Available from:
  \url{https://www.pnas.org/doi/abs/10.1073/pnas.0601266103}.

\bibitem{germannSchoolDismissalPandemic2019}
Germann TC, Gao H, Gambhir M, Plummer A, Biggerstaff M, Reed C, et~al.
\newblock School Dismissal as a Pandemic Influenza Response: {{When}}, Where
  and for How Long?
\newblock Epidemics. 2019 Sep;28:100348.
\newblock Available from:
  \url{https://www.sciencedirect.com/science/article/pii/S1755436518301749}.

\bibitem{delvalleEpiCastSimulatingEpidemics2021}
Del~Valle SY, Germann TC, Fairchild G, Manore CA, Smith MZ, Dauelsberg LR,
  et~al.
\newblock {{EpiCast}}: {{Simulating Epidemics}} with {{Extreme Detail}}.
\newblock Los Alamos National Laboratory (LANL), Los Alamos, NM (United
  States); 2021. LA-UR-21-24603.
\newblock Available from: \url{https://www.osti.gov/biblio/1783478}.

\bibitem{germannAssessingK12School2022}
Germann TC, Smith MZ, Dauelsberg LR, Fairchild G, Turton TL, Gorris ME, et~al.
\newblock Assessing {{K-12}} School Reopenings under Different {{COVID-19
  Spread}} Scenarios -- {{United States}}, School Year 2020/21: {{A}}
  Retrospective Modeling Study.
\newblock Epidemics. 2022 Dec;41:100632.
\newblock Available from:
  \url{https://www.sciencedirect.com/science/article/pii/S175543652200072X}.

\bibitem{chaoModelingLayeredNonpharmaceutical2020}
Chao DL, Oron AP, Srikrishna D, Famulare M. Modeling Layered Non-Pharmaceutical
  Interventions against {{SARS-CoV-2}} in the {{United States}} with
  {{Corvid}}. medRxiv; 2020.
\newblock Available from:
  \url{https://www.medrxiv.org/content/10.1101/2020.04.08.20058487v1}.

\bibitem{chaoFluTEPubliclyAvailable2010}
Chao DL, Halloran ME, Obenchain VJ, Longini IM.
\newblock {{FluTE}}, a {{Publicly Available Stochastic Influenza Epidemic
  Simulation Model}}.
\newblock PLOS Computational Biology. 2010 Jan;6(1):e1000656.
\newblock Available from:
  \url{https://journals.plos.org/ploscompbiol/article?id=10.1371/journal.pcbi.1000656}.

\bibitem{ozikPopulationDatadrivenWorkflow2021}
Ozik J, Wozniak JM, Collier N, Macal CM, Binois M.
\newblock A Population Data-Driven Workflow for {{COVID-19}} Modeling and
  Learning.
\newblock The International Journal of High Performance Computing Applications.
  2021 Sep;35(5):483-99.
\newblock Available from: \url{https://doi.org/10.1177/10943420211035164}.

\bibitem{barrettEpiSimdemicsEfficientAlgorithm2008}
Barrett CL, Bisset KR, Eubank SG, Feng X, Marathe MV.
\newblock {{EpiSimdemics}}: {{An}} Efficient Algorithm for Simulating the
  Spread of Infectious Disease over Large Realistic Social Networks.
\newblock In: {{SC}} '08: {{Proceedings}} of the 2008 {{ACM}}/{{IEEE
  Conference}} on {{Supercomputing}}; 2008. p. 1-12.
\newblock Available from: \url{https://ieeexplore.ieee.org/document/5214892}.

\bibitem{chenEpihiperHighPerformance2025}
Chen J, Hoops S, Mortveit HS, Lewis BL, Machi D, Bhattacharya P, et~al.
\newblock Epihiper---{{A}} High Performance Computational Modeling Framework to
  Support Epidemic Science.
\newblock PNAS Nexus. 2025 Jan;4(1):pgae557.
\newblock Available from: \url{https://doi.org/10.1093/pnasnexus/pgae557}.

\bibitem{bissetSimulatingSpreadInfectious2012}
Bisset KR, Aji AM, Bohm E, Kale LV, Kamal T, Marathe MV, et~al.
\newblock Simulating the {{Spread}} of {{Infectious Disease}} over {{Large
  Realistic Social Networks Using Charm}}++.
\newblock In: 2012 {{IEEE}} 26th {{International Parallel}} and {{Distributed
  Processing Symposium Workshops}} \& {{PhD Forum}}; 2012. p. 507-18.
\newblock Available from: \url{https://ieeexplore.ieee.org/document/6270685}.

\bibitem{fergusonReport9Impact2020a}
Ferguson N, Laydon D, Nedjati~Gilani G, Imai N, Ainslie K, Baguelin M, et~al.
\newblock Report 9: {{Impact}} of Non-Pharmaceutical Interventions ({{NPIs}})
  to Reduce {{COVID19}} Mortality and Healthcare Demand.
\newblock Imperial College London; 2020.
\newblock Available from:
  \url{http://spiral.imperial.ac.uk/handle/10044/1/77482}.

\bibitem{kerrCovasimAgentbasedModel2021}
Kerr CC, Stuart RM, Mistry D, Abeysuriya RG, Rosenfeld K, Hart GR, et~al.
\newblock Covasim: {{An}} Agent-Based Model of {{COVID-19}} Dynamics and
  Interventions.
\newblock PLOS Computational Biology. 2021 Jul;17(7):e1009149.
\newblock Available from:
  \url{https://journals.plos.org/ploscompbiol/article?id=10.1371/journal.pcbi.1009149}.

\bibitem{hinchOpenABMCovid19AnAgentbasedModel2021}
Hinch R, Probert WJM, Nurtay A, Kendall M, Wymant C, Hall M, et~al.
\newblock {{OpenABM-Covid19}}---{{An}} Agent-Based Model for Non-Pharmaceutical
  Interventions against {{COVID-19}} Including Contact Tracing.
\newblock PLOS Computational Biology. 2021 Jul;17(7):e1009146.
\newblock Available from:
  \url{https://journals.plos.org/ploscompbiol/article?id=10.1371/journal.pcbi.1009146}.

\bibitem{shattockImpactVaccinationNonpharmaceutical2022}
Shattock AJ, Le~Rutte EA, D{\"u}nner RP, Sen S, Kelly SL, Chitnis N, et~al.
\newblock Impact of Vaccination and Non-Pharmaceutical Interventions on
  {{SARS-CoV-2}} Dynamics in {{Switzerland}}.
\newblock Epidemics. 2022 Mar;38:100535.
\newblock Available from:
  \url{https://www.sciencedirect.com/science/article/pii/S1755436521000785}.

\bibitem{halloranContainingBioterroristSmallpox2002}
Halloran ME, Longini IM, Nizam A, Yang Y.
\newblock Containing {{Bioterrorist Smallpox}}.
\newblock Science. 2002 Nov;298(5597):1428-32.
\newblock Available from:
  \url{https://www.science.org/doi/10.1126/science.1074674}.

\bibitem{tuccilloUrbanPopSpatialMicrosimulation2023}
Tuccillo J, Stewart R, Rose A, Trombley N, Moehl J, Nagle N, et~al.
\newblock {{UrbanPop}}: {{A}} Spatial Microsimulation Framework for Exploring
  Demographic Influences on Human Dynamics.
\newblock Applied Geography. 2023 Feb;151:102844.
\newblock Available from:
  \url{https://www.sciencedirect.com/science/article/pii/S0143622822002156}.

\bibitem{lodes2019}
{{LEHD Origin-Destination Employment Statistics Data}} Version 7. US Census
  Bureau, Longitudinal-Employer Household Dynamics Program; 2019.
\newblock Available from: \url{https://lehd.ces.census.gov/data/#lodes}.

\bibitem{mansonNationalHistoricalGeographic2024}
Manson S, Schroeder J, Van~Riper D, Knowles K, Kugler T, Roberts F, et~al..
  National {{Historical Geographic Information System}}: {{Version}} 19.0.
  Minneapolis, MN: IPUMS; 2024.
\newblock Available from:
  \url{https://www.nhgis.org/geographic-crosswalks#download-2010-2020}.

\bibitem{bureauoftransportationstatisticsTranStats2019}
{Bureau of Transportation Statistics}. {{TranStats}}; 2019.
\newblock Available from: \url{https://www.transtats.bts.gov/}.

\bibitem{NytimesCovid19data2023}
Nytimes/Covid-19-Data. The New York Times; 2023.
\newblock Available from: \url{https://github.com/nytimes/covid-19-data}.

\bibitem{CountyBusinessPatterns2019}
County {{Business Patterns}}. US Census Bureau; 2019.
\newblock Available from:
  \url{https://www.census.gov/data/datasets/2019/econ/cbp/2019-cbp.html}.

\bibitem{lowery-northMeasuringSocialContacts2013}
{Lowery-North} DW, Hertzberg VS, Elon L, Cotsonis G, Hilton SA, Ii CFV, et~al.
\newblock Measuring {{Social Contacts}} in the {{Emergency Department}}.
\newblock PLOS ONE. 2013 Aug;8(8):e70854.
\newblock Available from:
  \url{https://journals.plos.org/plosone/article?id=10.1371/journal.pone.0070854}.

\bibitem{ONETOnline2024}
O*{{NET Online}}. National Center for O*NET Development; 2024.
\newblock Available from: \url{www.onetonline.org}.

\bibitem{avdiuWhenFacetofaceInteractions2020}
Avdiu B, Nayyar G.
\newblock When Face-to-Face Interactions Become an Occupational Hazard:
  {{Jobs}} in the Time of {{COVID-19}}.
\newblock Economics Letters. 2020 Dec;197:109648.
\newblock Available from:
  \url{https://www.sciencedirect.com/science/article/pii/S0165176520304080}.

\bibitem{ElementarySecondaryInformation2019}
Elementary and {{Secondary Information System}}. National Center for Education
  Statistics; 2019.
\newblock Available from: \url{https://nces.ed.gov/ccd/elsi/}.

\bibitem{NationalTeacherPrincipal2020}
National {{Teacher}} and {{Principal Survey}} ({{NTPS}}). National Center for
  Education Statistics; 2020/2021.
\newblock Available from:
  \url{https://nces.ed.gov/surveys/ntps/estable/table/ntps/ntps2021_sflt07_t1s}.

\bibitem{MPIMessagePassingInterface2015}
{{MPI}}: {{A Message-Passing Interface Standard Version}} 3.0; 2015.
\newblock Available from:
  \url{https://www.mpi-forum.org/docs/mpi-3.0/mpi30-report.pdf}.

\bibitem{quinnRacialDisparitiesExposure2011}
Quinn SC, Kumar S, Freimuth VS, Musa D, {Casteneda-Angarita} N, Kidwell K.
\newblock Racial {{Disparities}} in {{Exposure}}, {{Susceptibility}}, and
  {{Access}} to {{Health Care}} in the {{US H1N1 Influenza Pandemic}}.
\newblock American Journal of Public Health. 2011 Feb;101(2):285-93.
\newblock Available from:
  \url{https://ajph.aphapublications.org/doi/10.2105/AJPH.2009.188029}.

\bibitem{bassettVariationRacialEthnic2020}
Bassett MT, Chen JT, Krieger N.
\newblock Variation in Racial/Ethnic Disparities in {{COVID-19}} Mortality by
  Age in the {{United States}}: {{A}} Cross-Sectional Study.
\newblock PLOS Medicine. 2020 Oct;17(10):e1003402.
\newblock Available from:
  \url{https://journals.plos.org/plosmedicine/article?id=10.1371/journal.pmed.1003402}.

\bibitem{fergusonGeographicTemporalVariation2022}
Ferguson JM, Justice AC, Osborne TF, Magid HSA, Purnell AL, Rentsch CT.
\newblock Geographic and Temporal Variation in Racial and Ethnic Disparities in
  {{SARS-CoV-2}} Positivity between {{February}} 2020 and {{August}} 2021 in
  the {{United States}}.
\newblock Scientific Reports. 2022 Jan;12(1):273.
\newblock Available from:
  \url{https://www.nature.com/articles/s41598-021-03967-5}.

\bibitem{irizarEthnicInequalitiesCOVID192023}
Irizar P, Pan D, Kapadia D, B{\'e}cares L, Sze S, Taylor H, et~al.
\newblock Ethnic Inequalities in {{COVID-19}} Infection, Hospitalisation,
  Intensive Care Admission, and Death: A Global Systematic Review and
  Meta-Analysis of over 200 Million Study Participants.
\newblock eClinicalMedicine. 2023 Mar;57:101877.
\newblock Available from:
  \url{https://www.sciencedirect.com/science/article/pii/S2589537023000548}.


\end{thebibliography}

\end{document}